\documentclass[aps,prb,showpacs,twocolumn,superscriptaddress,amsmath,amssymb]{revtex4-1}
\usepackage{graphicx,color}
\usepackage{bm}

\newcommand{\abs}[1]{\ensuremath{|#1|}}
\newcommand{\braket}[2]{\ensuremath{\langle #1|#2\rangle}}
\newcommand{\braketop}[3]{\ensuremath{\langle #1|#2|#3\rangle}}
\newcommand{\bra}[1]{\ensuremath{\langle #1|}}
\newcommand{\ket}[1]{\ensuremath{|#1\rangle}}
\newcommand{\tr}{\ensuremath{\operatorname{tr}}}
\newcommand{\sgn}{\ensuremath{\operatorname{sgn}}}

\begin{document}
	\title{Double quantum dot Cooper-pair splitter at finite couplings}
	
	\author{Robert Hussein}
	\affiliation{SPIN-CNR, Via Dodecaneso 33, 16146 Genova, Italy}
	\author{Lina Jaurigue}
	\affiliation{School of Chemical and Physical Sciences and MacDiarmid Institute for Advanced Materials and Nanotechnology, 
	Victoria University of Wellington, P.O. Box 600, Wellington 6140, New Zealand}
	\author{Michele Governale}
	\affiliation{School of Chemical and Physical Sciences and MacDiarmid Institute for Advanced Materials and Nanotechnology, 
	Victoria University of Wellington, P.O. Box 600, Wellington 6140, New Zealand}
	\author{Alessandro Braggio}
	\affiliation{SPIN-CNR, Via Dodecaneso 33, 16146 Genova, Italy}
	\affiliation{INFN, Sez. Genova, Via Dodecaneso 33, 16146 Genova, Italy}
	\date{\today}
	
	\begin{abstract}
		We consider the sub-gap physics of a hybrid double-quantum dot Cooper-pair splitter with large single-level spacings, in the presence of tunnelling between the dots and finite Coulomb intra- and inter-dot Coulomb repulsion. In the limit of a large superconducting gap, we treat the coupling of the dots to the superconductor exactly. 
		We employ a generalized master-equation method which easily yields  currents, noise and cross-correlators. 
		In particular, for  finite inter- and intra-dot Coulomb interaction, we investigate how the transport properties are determined by the 
		interplay between local and nonlocal tunneling processes between the superconductor and the dots. We examine the effect of inter-dot tunneling on the particle-hole symmetry of the currents with and without spin-orbit interaction. We show that spin-orbit interaction in combination with finite Coulomb energy opens the 
		possibility to control the nonlocal entanglement and its symmetry  (singlet/triplet). We demonstrate that the generation of nonlocal entanglement can be achieved even without
		any %
		direct nonlocal coupling to the superconducting lead.
	\end{abstract}
	
	\maketitle

	\section{Introduction}
	Recent developments in quantum technologies\cite{MonroeNature2002a,MartinisQuantumIP2009a,LaddNature2010a} have shown an
	enormous potential for applications.
	Quantum key distributions in quantum cryptography\cite{LoNatureP2014a} have became almost a standard technology.
	This progress was mainly realized  in optical systems. In order to enable the full potential of quantum technologies, spintronics\cite{LinderNatPhys2015a} and topotronics,\cite{YoshimiNatureC2015a} in solid state systems, it is crucial to be able to generate entangled states. 
	A promising %
	route to entanglement generation is offered
	by hybrid superconducting nanostructures. This type of system has very reach physics. 
	For example, the possibility to emulate topological superconductors in low dimensions with, possibly, the creation of Majorana
	bound states has clearly shown a revolutionary potential.\cite{QiRMP2011a,KitaevPU2001a,NayakRMP2008a,AliceRPP2012a}
	The enormous advancement in the production and control of nanotubes and nanowires\cite{RomeoNanoL2012a,RossellaNatNano2014a} opened up the possibility to couple nanosystems, in a very controlled way, to
	superconductors\cite{GiazottoNatPhys2011a,RoddaroNanoR2011a,SpathisNanotechnology2011a}
	taking advantage of their properties such as the spin-orbit (SO) interaction. 
	Quantum phase transitions and anomalous current-phase relations have been studied in hybrid semiconductor-superconductor 
	devices.\cite{LeeNatNano2014a,YokoyamaPRB2014a,MironovPRL2015a,ZondaSR2015a,MarraPRB2016a}
	SO interaction in the presence of superconducting correlations may lead to the generation of triplet ordering 
	in nanowires\cite{ShekhterPRL2016a} or quantum wells.\cite{YuPRB2016a}
	
	Superconductors are a natural source of electron-singlets (Cooper-pairs) which may provide nonlocal entangled 
	electrons when split.\cite{BouchiatNanotechnology2003a,RussoPRL2005a,SchindelePRL2012a,FueloepPRB2014a,LinderNatPhys2015a} 
	Semiconductor-superconductor-hybrid devices have been the object of experimental studies investigating signatures of nonlocal transport in charge currents and cross-correlations.\cite{DasNatureC2012a,SchindelePRB2014a,HeNatureC2014a}
	
	Cooper-pair splitting has also  been investigated in Josephson junctions.\cite{IshizakaPRB1995a,DeaconNatureC2015a} 
	Spin entanglement\cite{RecherPRB2001a,SothmannPRB2014a} and electron transport\cite{HammerPRB2007a,ChevallierPRB2011a,RechPRB2012a,DrosteJP2012a,TrochaPRB2015a} in hybrid 
	systems have been theoretically studied using full counting statistics (FCS)\cite{BelzigEPL2003a,GovernalePRB2008a,MortenPRB2008a,DrosteJP2012a,FuttererPRB2013a,SollerEPL2014a,StegmannArXiv2016a}. 
	Further studies have investigated the effects of external magnetic fields\cite{FueloepPRL2015a} and thermal gradients\cite{MachonPRL2013a,CaoAPL2015a} on Cooper-pair splitting.
	Quantum dots increase the efficiency of Cooper-pair splitting since sufficiently large intradot Coulomb interaction suppresses local 
	Cooper-pair tunneling.\cite{RecherPRB2001a,HofstetterNature2009a,SchindelePRL2012a,SchindelePRB2014a,FueloepPRL2015a} Alternatively, 
	the efficiency can be improved using spin-filtering as in spin valves.\cite{BeckmannPRL2004a,SammJAP2014a}
	Typically all these systems are investigated assuming a very strong on-site Coulomb interaction.  In the present paper, we consider the possibility of a weak  Coulomb interaction which complicates the analysis as it introduces additional transport channels. 
	We find that this is not necessarily a limitation in the creation of nonlocal entanglement, instead it offers a different route to achieve nonlocal entanglement.
	
\begin{figure}[t]
		\includegraphics[width=0.9\columnwidth]{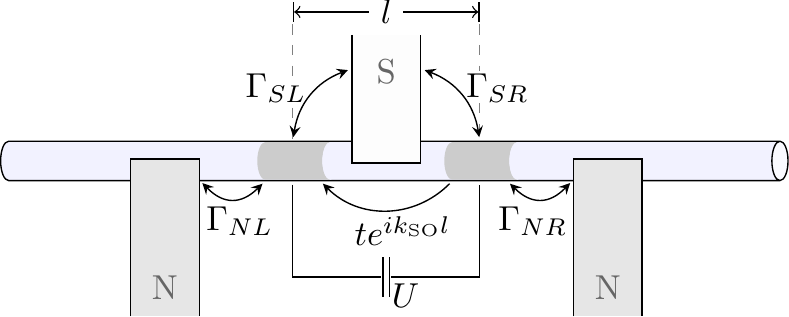}\caption{Double quantum dot circuit coupled to a $s$-wave superconductor
		acting as a Cooper-pair splitter. Electron-singlets nonlocally injected by the superconductor ($S$)
		into the double quantum dot can leave the system through opposite normal leads ($N$). Hence this system can be operated as a source of nonlocal entangled electron pairs.
		\label{fig.:model}}
	\end{figure}

	The model studied in this paper is a Cooper-pair splitter based on a double quantum dot (DQD) circuit that is tunnel coupled 
	to one superconductor and to two normal leads, see Fig~\ref{fig.:model}. This is an extension of the model studied by Eldridge et al., 
	Ref.~\onlinecite{EldridgePRB2010a}, to finite interdot tunneling and SO interaction.
	In this work, we investigate the effect of both local and nonlocal Cooper-pair tunneling on the 
	current and conductance in the presence of finite Coulomb energies.  Finally, we will discuss how 
	interdot tunneling with or without SO interaction affects the generation of nonlocal entanglement. 
	
	This work is organized as follows. In section~\ref{sec.:model} we introduce the model and the formalism employed for our calculations.
	In Section~\ref{sec.:transportOverview} we provide an overview of the transport properties in the absence of inter-dot tunneling. The effect of interdot tunneling and SO
	interaction is discussed in section \ref{sec.:so} . Finally, section~\ref{sec.:conclusions} is devoted to conclusions.
	
	\section{\label{sec.:model}Model and master-equation}
	\subsection{Model of the hybrid double quantum-dot system}
	The system under consideration, depicted schematically in Fig.~\ref{fig.:model}, consists of two quantum dots tunnel coupled to a common $s$-wave superconductor and each individually to a separate normal lead.\cite{SchindelePRL2012a,SchindelePRB2014a}
	The double-quantum-dot (DQD) system is modelled by the Hamiltonian
	\begin{align}
H_{\textrm{DQD}}= { }&
	 \sum_{\alpha,\sigma}\epsilon_{\alpha} n_{\alpha\sigma}
	+\sum_{\alpha}U_{\alpha} n_{\alpha\uparrow}n_{\alpha\downarrow}
	+U \sum_{\sigma, \sigma'} n_{L\sigma} n_{R\sigma'}
\nonumber\\
&+\Big(
	\frac{t}{2} \sum_{\sigma} e^{\sgn(\sigma)i\phi} d^\dag_{L\sigma}d_{R\sigma} +\mathrm{H.c.}
\Big), \label{eq.:HDQD}
	\end{align}
	where $\alpha=L,R$ labels the left and right dot, respectively, and $\sigma=\uparrow,\downarrow$ denotes the spin.
	The orbital levels  $\epsilon_{\alpha}$ are spin degenerate, and $U_{\alpha}$ and $U$ denote the intra- and interdot Coulomb interaction, respectively. 
	We define the number operator $n_{\alpha\sigma}=d_{\alpha\sigma}^{\dag}d_{\alpha\sigma}$,  where $d_{\alpha\sigma}$ is the annihilation operator for an electron with spin $\sigma$ in dot $\alpha$.
	The last term in Eq.~\eqref{eq.:HDQD} describes interdot tunneling through a barrier with SO coupling.\cite{DrosteJP2012a}  The phase $\phi$ is the phase acquired by a spin-up electron when tunneling from the right dot to the left one and it can be expressed as $\phi\equiv k_{\textrm{SO}}l\neq0$ where the SO strength is measured in terms of the wave number $k_{\textrm{SO}}$, $l$ is the interdot distance. Here, we used the convention $\sgn(\uparrow)=+1$ and $\sgn(\downarrow)=-1$.
	\footnote{Due to the absence of an applied magnetic field it is possible to choose the spin-quantization axis such that the interdot tunneling in the presence of the SO coupling is diagonal in the spin space.}
	The SO coupling may become relevant for InAs\cite{FasthPRL2007a,HernandezPRB2010a} 
	and InSb\cite{Nadj-PergePRL2012a,WeperenPRB2015a} nanowire devices where one finds 
	values of $1/k_{\text{SO}}$ of 
	typically 50-300 nm which are comparable to the typical distance between the two dots in these nanodevices.  The model Hamiltonian, Eq.~\eqref{eq.:HDQD}, gives an accurate description of the system when the single-particle level spacings in the dots are large compared to the other energy scales. In this limit, for $k_B T\ll U_\alpha$, at most 4 electrons can occupy the double-quantum-dot system. 
	
	The normal leads ($\eta=L,R$) are modeled as fermionic baths 
	while the superconducting lead ($\eta=S$) is described by the mean-field $s$-wave BCS Hamiltonian,
	\begin{equation}
		H_{\eta}=\sum_{k \sigma} \epsilon_{\eta k}
		c_{\eta k \sigma}^\dag c_{\eta k \sigma}- \delta_{\eta,S} \Delta\sum_{k}\big(  
		c_{\eta -k \downarrow} 
		c_{\eta k \uparrow}+\mathrm{H.c.}
		\big).
	\end{equation}
	Here, $c_{\eta k \sigma}$ ($c_{\eta k\sigma}^\dag$) are the fermionic annihilation (creation) operators of the leads
	and $\epsilon_{\eta k}$ are corresponding single-particle energies. 
	Without loss of generality the pair potential in the superconductor, $\Delta$, is
	chosen to be real and positive.  
	For convenience, we choose the chemical potential of the superconductor to be zero
	and use it as reference for the chemical potentials of the normal leads.
	
	The quantum dots are coupled to the normal leads and the superconductor
	via the Hamiltonian,
	$H_{\textrm{DQD-leads}}=\sum_{\eta\alpha}H^{\textrm{tunnel}}_{\eta\alpha}$,
	where the coupling of dot $\alpha$ with lead $\eta=L,R,S$ is described by the standard tunneling Hamiltonian 
	\begin{equation}
		H^{\textrm{tunnel}}_{\eta\alpha} =\sum_{k\sigma}\big( 
		V_{\eta\alpha}  c^{\dagger}_{\eta k\sigma}d_{\alpha\sigma} +\mathrm{H.c.}
		\big).
	\end{equation}
	Here, $V_{LR}=V_{RL}=0$ since the left (right) dot is not directly coupled to the right (left) lead. 
	The effective tunneling rates are $\Gamma_{\eta\alpha}=(2\pi/\hbar)\abs{V_{\eta\alpha}}^2\rho_{\eta}$
	where  the density of states $\rho_{\eta}$ in lead $\eta$ is assumed to be energy independent in the energy window relevant for the transport.
	For a better readability we introduce $\Gamma_{N\alpha}\equiv\Gamma_{\alpha\alpha}$ to emphasize the coupling to the normal leads with a subscript $N$.
	
	As we are interested in Cooper-pair splitting and in general sub-gap transport, we assume the superconducting gap to be the largest energy scale in the system. In this limit the quasi-particles in the superconductor are inaccessible and the superconducting 
	lead can be traced out exactly.\cite{RozhkovPRB2000a,MengPRB2009a,RajabiPRL2013a} Thus, the system dynamics 
	reduces to the effective Hamiltonian\cite{EldridgePRB2010a,BraggioSSC2011a,SothmannPRB2014a}
	\begin{align}
		\begin{split}
H_{S}=H_{\textrm{DQD}} & - \sum_{\alpha=L,R} \frac{\Gamma_{S\alpha}}{2} \big( 
	d^\dagger_{\alpha,\uparrow} d^\dagger_{\alpha,\downarrow}+\mathrm{H.c.}
\big)\\
& -\frac{\Gamma_S}{2}\big( 
	d^\dagger_{R,\uparrow} d^\dagger_{L,\downarrow}- d^\dagger_{R,\downarrow} d^\dagger_{L,\uparrow}+ \mathrm{H.c.}
\big)  \label{eq.:Heff}
		\end{split}
	\end{align}
	where $\Gamma_S$ describes the nonlocal  proximity effect. This nonlocal coupling decays with the interdot distance $l$, as 
	$\Gamma_S\sim\sqrt{\Gamma_{SL}\Gamma_{SR}}e^{-l/\xi}$, with $\xi$ being the coherence length of the Cooper-pairs.\cite{RecherPRB2001a}
	So, only values $0\leq \Gamma_S\leq\sqrt{\Gamma_{SL}\Gamma_{SR}}$ are physically admissible.
	The second term describes the local Andreev reflection (LAR) processes where  Cooper pairs tunnel locally from the superconductor 
	to dot $\alpha$. The last term describes cross-Andreev reflection (CAR), that  is a nonlocal 
	Cooper-pair tunneling process where  Cooper-pairs split into the two dots.  Due to CAR, electrons leaving the system through 
	opposite normal leads are potentially entangled. On the contrary, the LAR process does not contribute to the nonlocal entanglement production. The LAR process is  usually attenuated by large intradot couplings, $U_{\alpha}$. 
	
	Albeit the effective Hamiltonian, Eq.~\eqref{eq.:Heff}, no longer preserves the total particle number for the double-dot system
	it still preserves the parity of the total occupation, $\sum_{\alpha\sigma}n_{\alpha\sigma}$.
	A decomposition, $H_S=H_S^{\textrm{even}}\oplus H_S^{\textrm{odd}}$, of 
	the system Hamiltonian into an even and an odd parity sector is provided in Appendix~\ref{sec.:HS}. In conclusion the Hilbert space for the proximized double-dot system has the dimension 16. A generalization to include more charge states, to treat for instance smaller level spacings or higher temperatures,
	is straightforward and can be treated within the master-equation approach  presented below. 
	Lowest order corrections in $1/\Delta$ can be also included in the system Hamiltonian according to Ref.~\onlinecite{AmitaiPRB2016a}.
	
	In the following, we consider the case of the quantum dots weakly coupled to the normal leads in comparison to 
	the superconducting one, $\Gamma_{S\alpha}\gg\Gamma_{N\alpha'}$. 
	In this limit quantum transport is mainly characterized by the transitions between the eigenstates of $H_S$, 
	the Andreev bound states.\cite{EldridgePRB2010a,BraggioSSC2011a} Those tunneling events with the normal leads either add a single charge to the DQD 
	or remove one from it and, thus, change the parity of the DQD.

	\subsection{Master-equation and transport coefficients}
	We calculate the stationary transport properties, such as the current and the conductance, 
	by means of the master-equation formalism using standard FCS techniques.
	All the relevant transport properties can be related to the Taylor coefficients of the cumulant
	generating function\cite{BagretsPRB2003a,FlindtPRB2004a,BraggioPRL2006a,KaiserAP2007a,HusseinPRB2014a} and  obtained in an iterative scheme.\cite{FlindtPRL2008a,FlindtPRB2010a}
	In this work, we limit our analysis to the current and the differential conductance, however,
	also higher cumulants, such as noise and cross-correlations, can be easily obtained.
	
	Here, we consider the regime $\Gamma_{N\alpha}\ll k_BT$, for which the tunnel couplings to the 
	normal leads can be treated in first order.
	The tunnel couplings to the superconductor, the charging energies,
	and the interdot tunneling are treated exactly within the model under consideration.
	This leads to the master-equation $\dot P_a = \sum_{a'} \big( w_{a\leftarrow a'}P_{a'}- w_{a'\leftarrow a}P_{a}\big)$ for the occupation 
	probabilities $P_a$ of the eigenstates $\ket{a}$ of the system Hamiltonian, where $w_{a\leftarrow a'}$ 
	are Fermi golden rule rates. The tunneling rates for the tunneling-in contribution read
	\begin{align}
w_{a\leftarrow a'}^{\alpha\sigma, \textrm{in}}({\bm\chi})=e^{-i\chi_\alpha}
\Gamma_{N\alpha} f_\alpha(E_a-E_{a'})\abs{\braketop{a}{d_{\alpha\sigma}^\dag}{a'}}^2 \label{eq.:win}.
	\end{align}
	Here, $E_a$ and $\ket{a}$ refer to the eigenenergies and the eigenstates of $H_S$, and
	$f_\alpha(\epsilon)=\{1+\exp[(\epsilon-\mu_\alpha)/k_BT]\}^{-1}$ denotes the Fermi function of normal lead $\alpha$ with chemical potential $\mu_\alpha$ and temperature $T$.
	We only attach\cite{BagretsPRB2003a,BraggioPRL2006a} counting variables to the normal leads,
	${\bm\chi}=(\chi_L,\chi_R)$. The stationary current through the superconductor $I_S$,
	can be easily expressed in terms of the currents through the left and right leads, $I_S=-I_L-I_R$.
	The tunneling-out contribution can be obtained from the substitution 
	$\{d_{\alpha\sigma}^\dag,f_\alpha(\epsilon),\chi_\alpha\}\to\{d_{\alpha\sigma},\bar f_\alpha(-\epsilon),-\chi_\alpha\}$,
	where $\bar f_\alpha(\epsilon)=1-f_\alpha(\epsilon)$.
	Summation over the spin and lead indices yields the full rates $w_{a\leftarrow a'}=\sum_{\alpha\sigma}(
	w_{a\leftarrow a'}^{\alpha\sigma, \textrm{in}}+w_{a\leftarrow a'}^{\alpha\sigma, \textrm{out}})$. 
	
	Single electron tunneling changes the parity of the system. 
	So, the only transitions that occur are  between the 
	eigenstates $\ket{e_i}$ of $H_S$  with even occupation number and those with odd occupation numbers, $\ket{o_j}$. 
	\begin{table}[ht]
		\begin{ruledtabular}
			\begin{tabular}{lll}
				$\ket{0}$ & & empty state \\
				$\ket{S}$ & $=\frac{1}{\sqrt{2}} \big( 
				d^\dag_{R\uparrow} d^\dag_{L\downarrow} 
				-d^\dag_{R\downarrow} d^\dag_{L\uparrow}
				\big)\ket{0}$ & singlet state\\
				$\ket{d\alpha}$ & $= d_{\alpha\uparrow}^\dag  d_{\alpha\downarrow}^\dag \ket{0}$ & doubly occupied states\\
				$\ket{dd}$ & $=d_{R\uparrow}^\dag  d_{R\downarrow}^\dag d_{L\uparrow}^\dag  d_{L\downarrow}^\dag \ket{0}$
				& quadruply occupied state\\
				$\ket{T0}$ & $=\frac{1}{\sqrt{2}}\big( 
				d^\dag_{R\uparrow} d^\dag_{L\downarrow}
				+d^\dag_{R\downarrow} d^\dag_{L\uparrow}
				\big)\ket{0}$ & unpolarized triplet state\\
				$\ket{T\sigma}$ & $=d^\dag_{R\sigma} d^\dag_{L\sigma}\ket{0}$ & polarized triplet states\\
				\hline
				$\ket{\alpha\sigma}$ & $=d^\dag_{\alpha\sigma}\ket{0}$ & singly occupied states\\
				$\ket{t\alpha\sigma}$ & $=d^\dag_{\alpha\sigma}d^\dag_{\bar{\alpha}\uparrow}d^\dag_{\bar{\alpha}\downarrow}\ket{0}$
				& triply occupied states
			\end{tabular}
		\end{ruledtabular}
		\caption{\label{tab.:HS_basis} Choice of the system basis, subdivided into states with even parity (top cell) 
		and states with odd parity (bottom cell). Note that the indices $\alpha$ and $\sigma$ in $|\alpha\sigma\rangle$ and $|t\alpha\sigma\rangle$ refer 
		to the singly-occupied unpaired electron.}
	\end{table}
	Here, the indices $i,j=1\ldots8$ label the eigenstates of the even and odd parity sector, respectively. 
	We can write the eigenstates of the even sector in the basis of Table~\ref{tab.:HS_basis},
	\begin{align}
		\begin{split}
\ket{e_i} =	{ }&  e_{i,0}\ket{0} +e_{i,S}\ket{S} +\sum_\alpha e_{i,d\alpha}\ket{d\alpha}\\
			+& e_{i,dd}\ket{dd} +e_{i,T0}\ket{T0} +\sum_\sigma e_{i,T\sigma} \ket{T\sigma}. 
		\end{split}
	\end{align}
	Similarly the eigenstates of the odd sector can be expressed as
	\begin{align}
\ket{o_j} =	\sum_{\alpha \sigma}\big( o_{j,\alpha\sigma}\ket{\alpha\sigma}
			+o_{j,t\alpha\sigma}\ket{t\alpha\sigma}\big).
	\end{align}
	Finally we can evaluate the matrix elements of the fermionic operators, 
	$\braketop{o_j}{d_{\alpha\sigma}^{(\dag)}}{e_i}$, and therewith express the transitions from the state $\ket{e_i}$ to the  
	state $\ket{o_j}$ as
	\begin{align}
w_{o_j\leftarrow e_i} = &\sum_{\alpha\sigma}\Gamma_{N\alpha}\bar f_{\alpha}(E_{e_i}-E_{o_j})e^{i \chi_\alpha}\Big| 
	o_{j,\alpha\bar{\sigma}}^* e_{i,d\alpha}+o_{j,t\alpha\bar{\sigma}}^*e_{i,dd} \nonumber\\
&\quad+\frac{1}{\sqrt{2}}o_{j,\bar{\alpha}\bar{\sigma}}^*(e_{i,S}-\alpha\sigma e_{i,T0})
	-\alpha\sigma o^*_{j,\bar{\alpha}\sigma}e_{i,T\sigma}
\Big|^2 \nonumber\\
+&\sum_{\alpha\sigma}\Gamma_{N\alpha} f_{\alpha}(E_{o_j}-E_{e_i})e^{-i \chi_\alpha}\Big| 
	o_{j,\alpha\sigma}^* e_{i,0}+o_{j,t\alpha{\sigma}}^*e_{i,d\bar{\alpha}}\nonumber\\
&\quad-\frac{1}{\sqrt{2}}o_{j,t \bar{\alpha}{\sigma}}^*(e_{i,S}+\alpha\sigma e_{i,T0})
	-\alpha\sigma o^*_{j,t\bar{\alpha}\bar\sigma}e_{i,T\bar{\sigma}}
\Big|^2 \label{eq.:rateba},
	\end{align}
	with the coefficients $e_{i,a}=\braket{a}{e_i}$ and $o_{j,a}=\braket{a}{o_j}$ and $E_{e_i}$ ($E_{o_j}$) the eigenenergy corresponding to $\ket{e_i}$ ($\ket{o_j} $). The bar on
	the indices indicates their complement, i.e. $\bar L = R$, $\bar \uparrow = \downarrow$ and so forth.
	The rate for the inverse transition $w_{e_i\leftarrow o_j}$ follows straightforwardly.
	\begin{figure*}[t]
		\includegraphics[width=1.5\columnwidth]{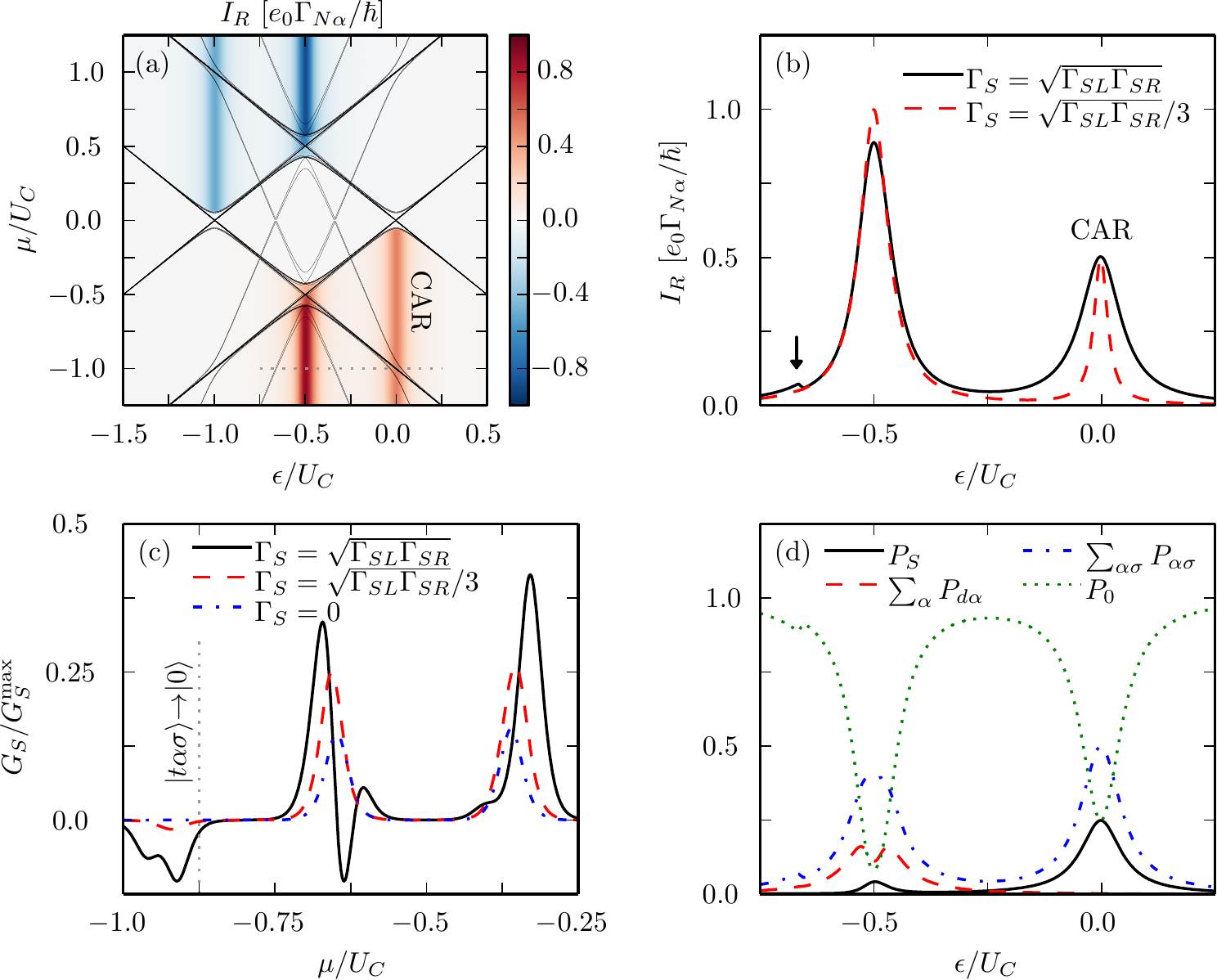}\caption{\label{fig.:limit_GSBiggerU}%
		(a) Current $I_R$ through the right lead as a function of the dots' level positions $\epsilon\equiv\epsilon_L=\epsilon_R$, and 
		the chemical potential $\mu\equiv\mu_L=\mu_R$, where the solid lines indicate the condition under which the 
		chemical potential is equal to the Andreev addition energies. Parameters are 
		$\Gamma_S=\Gamma_{S\alpha}=7.5\ 10^{-2}U_C$, $t=U=0$, and $k_BT=2.5\ 10^{-3}U_C$, and $\Gamma_{N\alpha}=2.5\ 10^{-4}U_C$.
		(b) Current $I_R$ as a function of the dots' level positions $\epsilon\equiv\epsilon_L=\epsilon_R$ at constant $\mu=-U_C$ for intermediate nonlocal coupling $\Gamma_S=\Gamma_{S\alpha}/3$ (dashed line),
		and maximal nonlocal coupling $\Gamma_S=\Gamma_{S\alpha}$ (solid line). The value $\mu=-U_C$ is indicated in panel (a) by a dotted line.
		The arrow indicates the transition $\ket{t\alpha\sigma}\to\ket{0}$.
		(c) Differential conductance $G_S\equiv dI_S/dV$ at constant $\epsilon=-0.625 U_C$
		for $\Gamma_{S\alpha}=10 k_BT=0.125 U_C$, normalized by its maximum $G_S^{\textrm{max}}$ at constant $\epsilon=-U_C/2$. (d) Occupation probabilities as a function of the level position for $\mu=-U_C$ and maximal nonlocal coupling  $\Gamma_S=\Gamma_{S\alpha}$. 
		}
	\end{figure*}
	
	From the cumulant generating function, 
	$Z(\bm\chi,\bm\mu)=\lim_{t\to\infty}\frac{\partial}{\partial t}\ln\tr P({\bm\chi,\bm\mu})$,
	we can obtain the stationary current $I_\alpha=\partial Z/\partial i\chi_\alpha|_{{\bm\chi}={\bm0}}$ in the normal lead $\alpha=L,R$ 
	and the corresponding differential conductance $G_{\alpha,\beta}=-\partial^2 Z/\partial i\chi_\alpha\partial\mu_\beta|_{{\bm\chi}={\bm0}}$.
	Both the current\cite{FlindtPRL2008a,FlindtPRB2010a} and the conductance\cite{HusseinPRB2014a} can be
	calculated in the usual iterative scheme.

	\section{\label{sec.:transportOverview}transport in absence of interdot tunneling}
	In this section, we give an overview of how the local and nonlocal proximization affects  quantum transport
	in absence of interdot tunneling. In particular, we start discussing the two limits $\Gamma_{S\alpha}\gg U$ and $U\gtrsim\Gamma_{S\alpha}$.
	Both limits feature a resonant current originating from the CAR process. The former case of weak interdot
	Coulomb energy is typically realized in experiments.\cite{SchindelePRL2012a}
	The limit of strong interdot Coulomb energy 
	additionally permits to study 
	resonant currents which are entirely characterized by LAR.
	Throughout this work, we consider identical quantum dots, i.e. $ U_L=U_R\equiv U_C$, 
	$\Gamma_{SL}=\Gamma_{SR}\equiv\Gamma_{S\alpha}$, and $\Gamma_N\equiv\Gamma_{NL}=\Gamma_{NR}$. We limit to the case of equal orbital levels in the two quantum dots, $\epsilon\equiv\epsilon_L{=}\epsilon_R$, and equal chemical potentials  $\mu\equiv \mu_L{=}\mu_R$. 
	
	\begin{figure*}[t]
		\includegraphics[width=1.5\columnwidth]{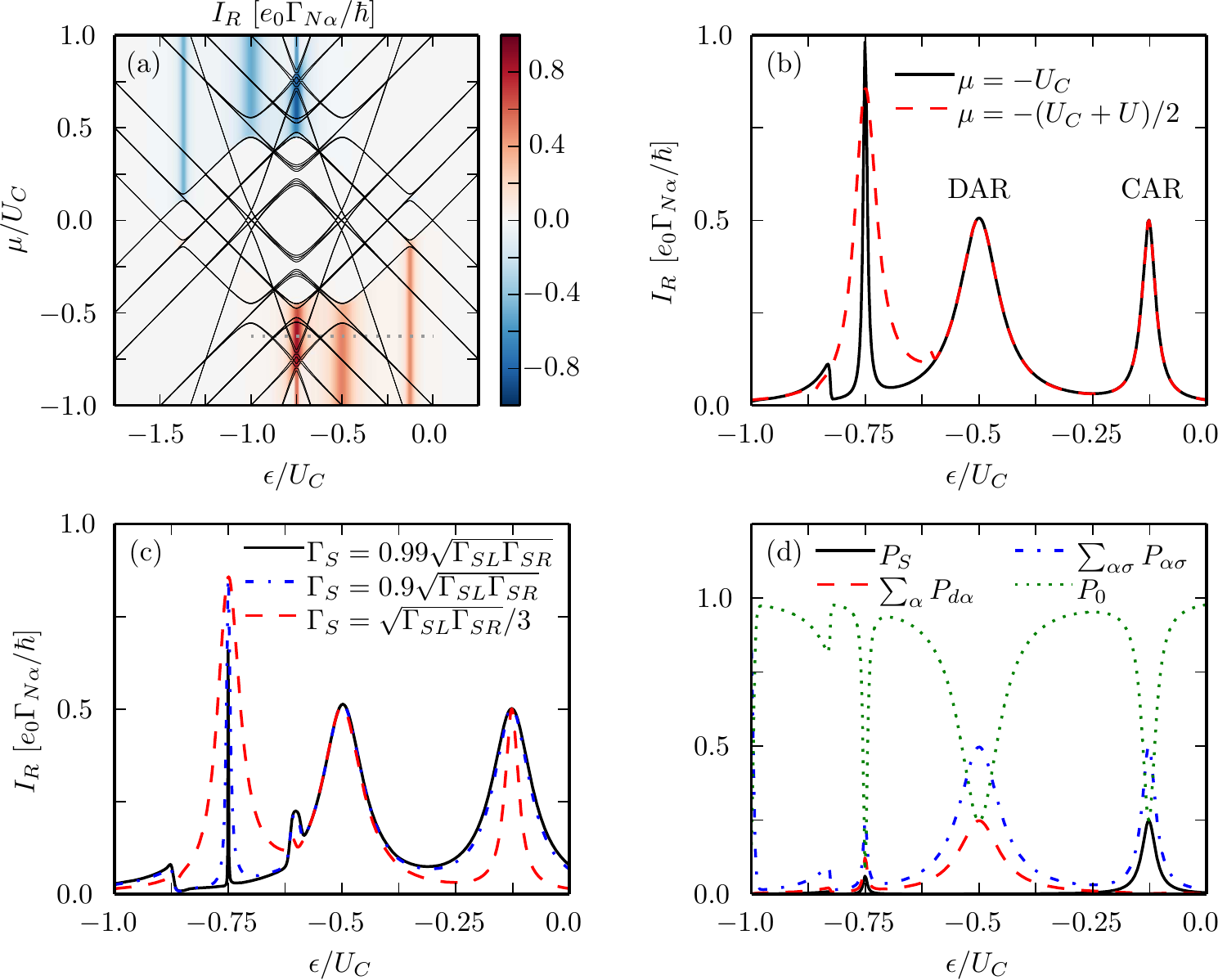}\caption{\label{fig.:limit_UApproxGS}%
		(a) Current $I_R$ through the right lead as a function of the gate voltage $\epsilon\equiv\epsilon_L=\epsilon_R$, and 
		the chemical potential $\mu\equiv\mu_L=\mu_R$ for finite $U=0.25\  U_C$ and $\Gamma_S=\Gamma_{S\alpha}/3$.
		Other parameters are as in Fig.~\ref{fig.:limit_GSBiggerU}. (b) Corresponding slices at constant $\mu=-(U_C+U)/2$ (dashed line), 
		and $\mu=-U_C$ (solid line). The slice at $\mu=-(U_C+U)/2$ is indicated in panel (a) by a dotted line.
		(c) Current $I_R$ at constant $\mu=-(U_C+U)/2$ for various values of $\Gamma_S$.
		(d) Corresponding population probabilities as a function of the  gate voltage $\epsilon$
		at fixed $\mu=-U_C$. 
		}
	\end{figure*}
	
	\subsection{Weak interdot Coulomb energy, $U\approx0$}
	
	In Fig.~\ref{fig.:limit_GSBiggerU}(a), we show a density plot of the current in the right lead $I_R(\epsilon,\mu) $ as a function of the dots' level $\epsilon$, which can be tuned by gate voltages, and the chemical potential $\mu$ of the normal leads. 
	We notice that the current in the normal leads obeys the symmetry $I_\alpha(\epsilon,\mu)=-I_\alpha(2\epsilon_0-\epsilon,-\mu)$ with $\epsilon_0=-(U_C/2+U)$. This symmetry is due the particle-hole (PH) symmetry of the Hamiltonian Eq.~\eqref{eq.:Heff} in the absence of interdot tunneling.
	
	We focus on the situation $\mu<0$ which corresponds to the transport of Cooper pairs from the superconductor to the double-dot system. 
	Two resonances can be seen in Fig.~\ref{fig.:limit_GSBiggerU}(a): one at $\epsilon=\epsilon_{\textrm{CAR}}=-U/2$ that is caused only by  CAR  and another at  
	$\epsilon=\epsilon_0$ which originates from both CAR and LAR. The current is asymmetrical with respect 
	to the chemical potentials $\mu$. The bias asymmetry of the CAR peak is due to the triplet blockade: for $\mu>0$ tunneling of electrons from the leads can bring the double-dot in a triplet state whose spin symmetry is incompatible with the BCS superconductor, hence blocking the CAR.\cite{EldridgePRB2010a}
	
	Along the level position axis, the CAR resonance is centered at  $\epsilon_{\textrm{CAR}}$ and its broadening is  $\sqrt{2}\Gamma_S$. 
	This can be seen in panel (b) of  Fig.~(\ref{fig.:limit_GSBiggerU}), which shows the current
	at constant $\mu=-U_C$ for two different values of the nonlocal coupling: $\Gamma_S=\sqrt{\Gamma_{SL}\Gamma_{SR}}/3$ (dashed line) and 
	$\Gamma_S=\sqrt{\Gamma_{SL}\Gamma_{SR}}$ (solid line). 
	The CAR broadening is proportional to the nonlocal coupling $\Gamma_S$ but its height does not depend on it. 
	The CAR resonance instead follows the singlet population, i.e. 
	$\hbar I_R^{\textrm{CAR}}/e_0\Gamma_{NR}\approx2P_S$, as can be seen in panel (d). 
	The states involved in the CAR process are $\ket{0}$, $\ket{\alpha\sigma}$, $\ket{S}$ and in fact one only observes the corresponding populations,
	$P_0+\sum_{\alpha\sigma}P_{\alpha\sigma}+P_{S}\approx1$.
	On the contrary, the resonance at $\epsilon=-U_C/2$ is mainly due to the LAR, but involves 
	also CAR as indicated by the non-vanishing singlet population at the LAR resonance [see panel (d)].
	
	We will discuss now that strong superconducting coupling may also generate negative differential conductance (NDC) when single electron tunneling events with the normal leads are accompanied by a simultaneous exchange of a Cooper-pair.
	For instance if one of the dots is doubly occupied, while the other is singly occupied, it can occur that an electron leaves the system through a normal-metal lead 
	and the two remaining electrons tunnel (locally or nonlocally) to the superconductor. If the process is energetically admissible the total current is reduced instead of increased by the opening of the new resonance and NDC is observed.  
	
	This is indeed observed in panel (c) of Fig.~\ref{fig.:limit_GSBiggerU} having defined the conductance as $G_S\equiv d I_S/dV$.
	We show $G_S$ as function of the 
	chemical potential, for a fixed level position $\epsilon=\epsilon_0-0.125 U_C$.
	The differential conductance becomes negative around $\mu\approx3\epsilon + U_C$ (leftmost peak).
	This extra resonance corresponds energetically to the transition from the triply occupied states to the empty state, $\ket{t\alpha\sigma}\to\ket{0}$, where two electrons tunnel in the superconductor and the remaining electron tunnels in one of the normal leads. This involves \emph{only} the exchange of a nonlocal Cooper pair and is no longer present in the absence of nonlocal coupling [dot-dashed line in panel (c) of Fig.~\ref{fig.:limit_GSBiggerU}]. 
	In order to increase the visibility of the NDC we have chosen a stronger nonlocal coupling $\Gamma_{S}$ (by increasing $\Gamma_{S\alpha}$) to obtain a higher peak value and slightly higher temperatures to increase the linewidth of this resonance in comparison to other figures.

	\subsection{Finite interdot Coulomb energy, $U\gtrsim\Gamma_{S\alpha}$}%
	For finite interdot Coulomb energy the LAR dominated resonance in Fig.~\ref{fig.:limit_GSBiggerU}(a)
	splits into two resonances at gate voltages $\epsilon_{\textrm{LAR}}=-U_C/2$
	and $\epsilon_{0}=\epsilon_{\textrm{LAR}}-U$ as can be seen in panel (a) of Fig.~\ref{fig.:limit_UApproxGS}.
	The former current resonance is purely affected by LAR involving the states 
	$\ket{0}$, $\ket{\alpha\sigma}$, $\ket{d\alpha}$. In fact, in panel~\ref{fig.:limit_UApproxGS}(d) one observes 
	that only the corresponding populations are non vanishing, i.e. $P_0+\sum_{\alpha\sigma}P_{\alpha\sigma}+\sum_{\alpha}P_{d\alpha}\approx1$ and that 
	the current is proportional the population of the doubly occupied state, $\hbar I_R^{\textrm{LAR}}/e_0\Gamma_{NR}\approx4P_{dR}$. 
	We still observe the asymmetry of the nonlocal-current resonances in the chemical potential due to the triplet blockade.
	Additionally, one notices an asymmetry of the LAR resonances, which can be explained partly by the triplet blockade mechanism and 
	partly by energy considerations. 
	
	The central resonance at $\epsilon_{0}=\epsilon_{\textrm{LAR}}-U$ is affected both by the LAR and 
	the CAR processes. For an intermediate value of the chemical potentials [see dashed line in panel (b) of  Fig.~(\ref{fig.:limit_UApproxGS})] its width is roughly 
	proportional to $\sqrt{\Gamma_{SL}\Gamma_{SR}}-\Gamma_S$ and vanishes if $\Gamma_S$ becomes maximal. %
	In panel (c) we demonstrate the effect of the nonlocal Cooper-pair tunneling on the current resonances, for an intermediate value of the chemical potentials and for different values of the nonlocal coupling $\Gamma_S$. The dashed line corresponds to the dashed line in panel (b). When $\Gamma_{S}$ approaches its maximum, the width of the 
	central resonance (left peak) tends to zero 
	while its height remains unaffected.
	We suspect that the behavior of the width of this central resonance is due to the mutual exclusion
	of the local Cooper-pair tunneling process and the nonlocal one, and originates from a
	destructive interference of the two channels (see also later). On the contrary, if both processes were independent,  the linewidth would be the sum of both contributions. 
	In conclusion this regime of finite interdot Coulomb energy can be helpful to assess  the strength of the nonlocal coupling $\Gamma_S$ in comparison to the local terms. 
	
	\section{\label{sec.:so}Influence of interdot tunneling and spin-orbit interaction}
	\begin{figure}[t]
		\includegraphics{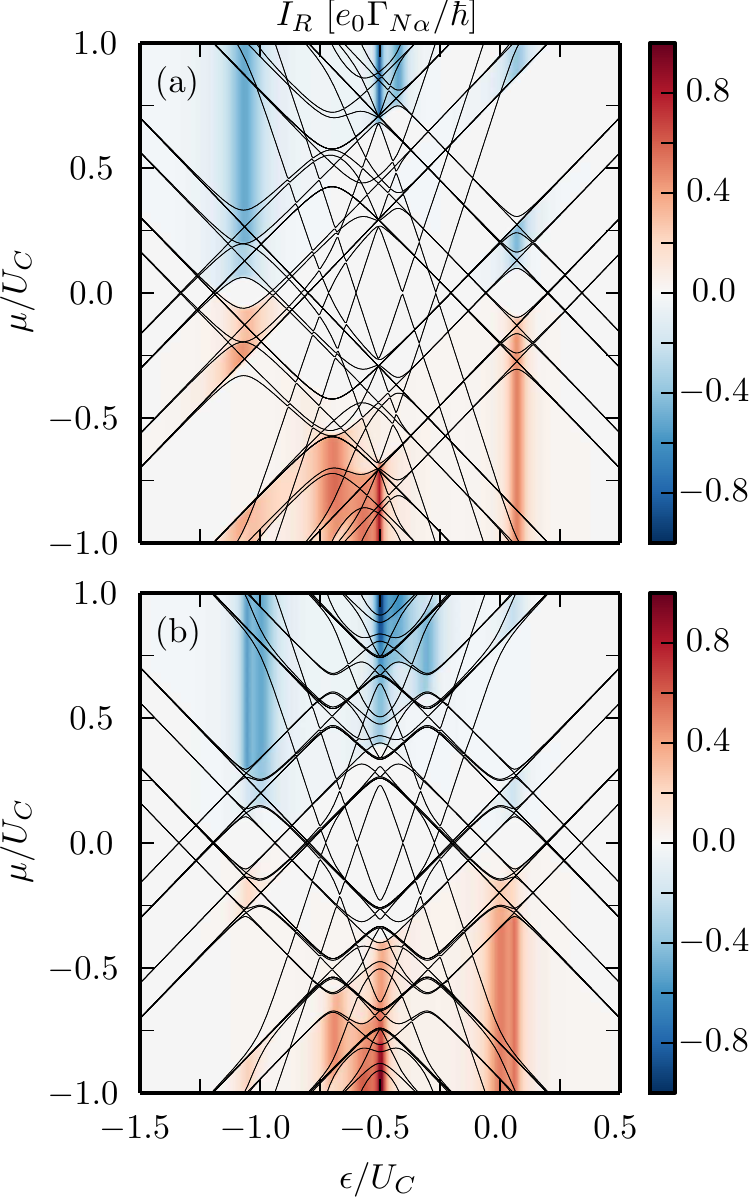}\caption{\label{fig.:finite_so}%
		(a) Current $I_R$ through the right lead as a function  of the gate voltage $\epsilon=\epsilon_L=\epsilon_R$ and 
		chemical potential $\mu=\mu_R=\mu_L$ for finite interdot tunneling with $t=0.4\: U_C$ and the SO angle $\phi=0$. 
		Other parameters are $U=0$, $\Gamma_S=\Gamma_{S\alpha}=7.5\ 10^{-2}U_C$, $\Gamma_{N\alpha}=2.5\ 10^{-4}U_C$ and $k_BT=2.5\ 10^{-3}U_C$. 
		(b) Current $I_R$ for the same parameters as in panel (a) but for finite SO interaction with an SO angle of $\phi=\pm\pi/2$.
		}
	\end{figure}
	
	In this section, we consider the effect of finite interdot tunneling and SO
	interaction on the current $I_R$. 
	For the sake of simplicity we consider in the following only the case $\phi=0$ (no SO coupling) and $\phi=\pm \pi/2$ (finite SO coupling with $k_{SO}l=\pi/2$).  
	Let us first focus on the general behavior of the current as a function of the level position and chemical potential as 
	shown in the density plots of  Fig. \ref{fig.:finite_so}. For simplicity we consider the case without interdot 
	Coulomb energy, $U=0$, which describes well the situation of $\Gamma_{S\alpha}\gg U$.   
	Finally, in order to see stronger signatures of the interdot tunneling term we generally 
	consider $U_C\gg t\gg\Gamma_{S},\Gamma_{S\alpha}$.
	
	In the top panel we show the case of finite interdot tunneling in the absence of SO coupling, i.e. $\phi=0$, which can be directly compared with the density plot of  Fig.~\ref{fig.:limit_GSBiggerU}(a) where the  interdot tunneling was absent. One immediately sees that the Andreev 
	resonant lines (black solid lines) are generally split in comparison to the case without interdot tunneling, giving rise to  an even richer Andreev-bound-state spectrum. 
	The most general observation is that the PH symmetry of the transport properties, as discussed in section \ref{sec.:transportOverview}, is broken, i.e. $I_\alpha(\epsilon,\mu)\neq-I_\alpha(2\epsilon_0-\epsilon,-\mu)$ with $\epsilon_0=-(U_C/2+U)$. 
	The breaking of the PH symmetry in transport is observed if both the quantities $\Gamma_S,\Gamma_{S\alpha}\neq 0$. On the other hand if one of these quantities vanishes the PH symmetry is restored. 
	We discuss PH-symmetry breaking in more detail in Sec.~\ref{sec.:breakingPH}.
	\begin{figure*}[t]
		\includegraphics{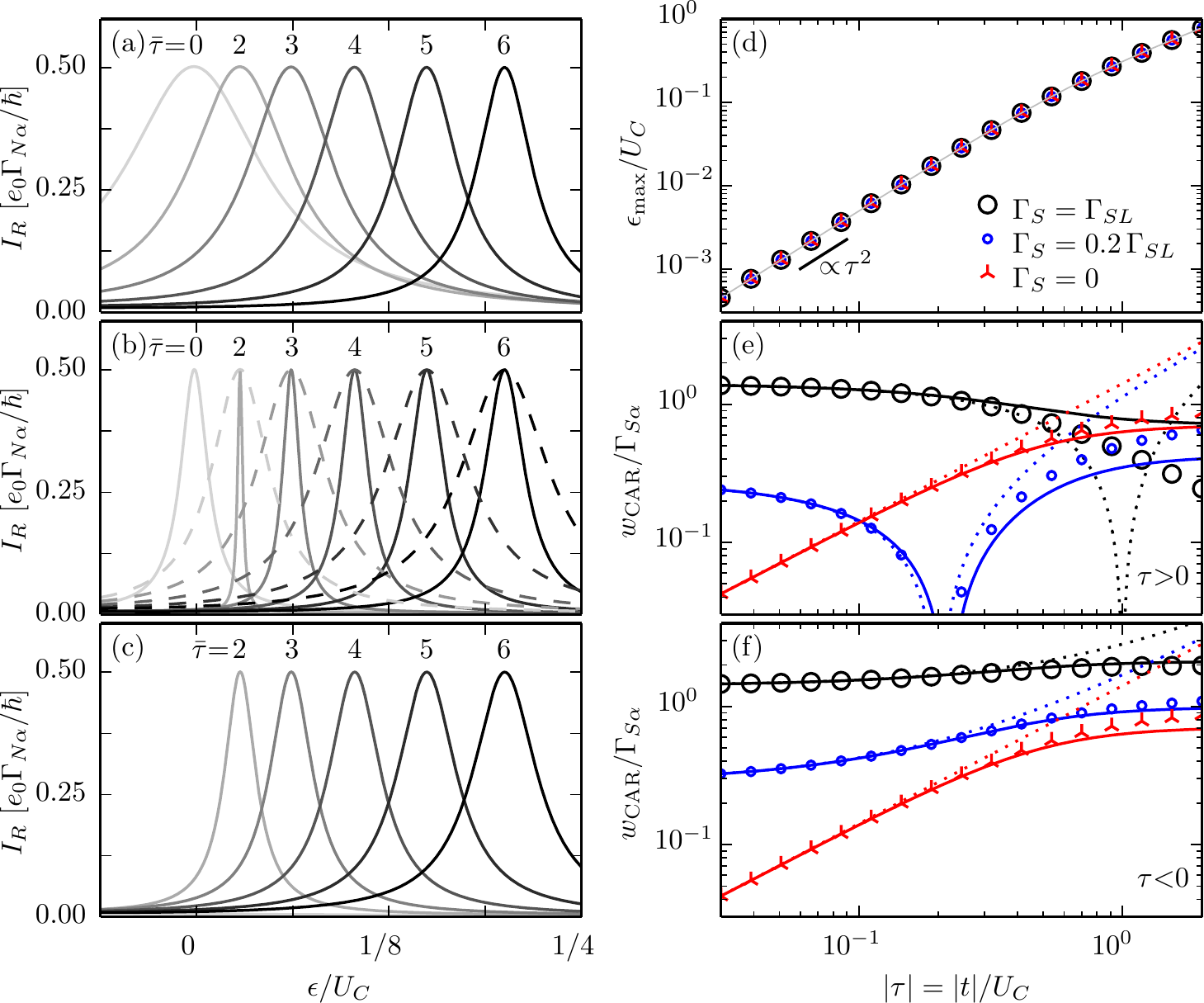}\caption{\label{fig.:CARpeak}%
		(a)-(c) Current $I_R$ in the CAR resonance as a function of the gate voltage for different values of the tunneling
		amplitude, $\bar\tau=8\abs{t}/U_C$, for the nonlocal term $\Gamma_S/\sqrt{\Gamma_{SR}\Gamma_{SL}}=1$ 
		[panel (a)], $1/3$ [panel (b)], $0$ [panel (c)] and other parameters as in Fig.~\ref{fig.:limit_GSBiggerU}.
		The solid lines correspond to $t/U_C>0$ while the dashed lines in panel (b) correspond to $t/U_C<0$.
		(d) Gate position at the maximum of the CAR resonance, $\epsilon_{\textrm{CAR}}$, as a function 
		of the scaling variable $\tau=t/U_C$ for the different values of the nonlocal term considered in the panels (a)-(c). (e)-(f) CAR resonance linewidth $w_{\mathrm{CAR}}$ as a function of the scaling variable $\abs{\tau}$ [$\tau>0$ for (e) and $\tau<0$ for (f)] for the different values of the nonlocal 
		term considered in the panels (a)-(c); the solid and dotted lines in panel (d)-(f) are the theoretical predictions in the simplified model discussed 
		in the main text.
		}
	\end{figure*}
	
	In the bottom panel of Figure \ref{fig.:finite_so}, instead, we show how the current is affected by tunneling in the presence of the SO coupling, for the case  $\phi=\pm\pi/2$.
	We see that in this case the PH symmetry is again restored for any value of  
	of $\Gamma_S$ and $\Gamma_{S\alpha}$. Note that the Andreev addition energies spectrum becomes also quite intricate and it is not so useful 
	to enter in the details of the behavior of any resonant line. In general one can see that in comparison to the 
	top panel crossings and avoided crossings occur between different pairs of Andreev levels. This is a natural consequence of the different 
	symmetry of the tunnel coupling between the two dots in the two cases. Finally, for $\Gamma_S,\Gamma_{S\alpha}\neq 0$,  
	the CAR peaks are split along the level-position axis and an extra resonance appears. 
	We will discuss in detail the nature of this extra resonance in Sec.~\ref{sec.:CARshift}
	
	\subsection{\label{sec.:breakingPH} Interdot tunneling and breaking of PH}

	To investigate the PH symmetry breaking, we apply the PH transformation $d_{\alpha\sigma}\to d^\dagger_{\alpha-\sigma}$ to Eq.~(\ref{eq.:Heff}).  
	It is easy to check that indeed this transformation leaves obviously unaffected the local and nonlocal 
	pairing  terms but is equivalent to a change of sign of the interdot tunneling term, i.e. $t\to-t$. Therefore, the symmetry obeyed by the current is $ I_\alpha(\epsilon,\mu,t) \to - I_\alpha(2\epsilon_0-\epsilon,-\mu,-t)$ which we have numerically verified.
	Notice that the sign of $t$ in the  tunneling Hamiltonian \emph{cannot} be gauged away only if \emph{both} the local and nonlocal pairing terms are present in Eq.~(\ref{eq.:Heff}). 
	Finally, we notice that for $\abs{\phi}=\pi/2$ the sign of $t$ is unessential 
	due to Kramer's degeneracy and therefore the PH symmetry is restored in this special case.
	
	It remains the question why the sign and more generally a phase of $t$ is detectable in the transport properties of the system. This is essentially due to the interference between two paths connecting the empty state with the singlet state. One path is the nonlocal Andreev tunneling with rate $\Gamma_S$ while the other is the process where a Cooper-pair virtually tunnels into one of the dots bringing it in the doubly occupied state and subsequently this state is converted into a singlet state by interdot tunneling. The interference between the two paths is clearly affected by the phase (not only the sign) of $t$. In order to observe this interference effect the doubly occupied state of a single dot needs to be accessible. We have verified that, for $U_C\to\infty$, an overall phase of $t$ does not affect the transport properties of the system.

	\subsection{\label{sec.:CARshift}Weak interdot Coulomb energy, $\Gamma_{S\alpha}\gg U$}
	We focus  on the effect of the interdot tunneling on  the CAR resonance.  In Fig.~\ref{fig.:CARpeak} (a)-(c) we show the evolution of the CAR current peak for different values of  $|t|$ for $\phi=0$. For increasing strength of the interdot tunneling, the position of the CAR resonance shifts to the  right  and at the same time the resonance linewidth changes. The peak shift is 
	$\delta \epsilon_{\textrm{CAR}}/U_C\approx(1/2)(t/U_C)^2$ for $t\ll U_C$. 
	This is shown 
	in the panel Fig.~\ref{fig.:CARpeak}(d) where the position of the CAR peak maximum  $\epsilon_{\textrm{max}}$ is plotted as a function of  $\tau=t/U_C$. For different values of the nonlocal coupling $\Gamma_S$ [different point styles 
	in panel (d)] the peak position follows the same universal function of $\tau$ (solid line). 
	Instead the linewidths, shown 
	in Fig.~\ref{fig.:CARpeak}(e)-(f), exhibit quite different behaviors depending on the value of  $\Gamma_S$, 
	the strength of $t$ and also its sign.

	\begin{figure}[t]
		\includegraphics{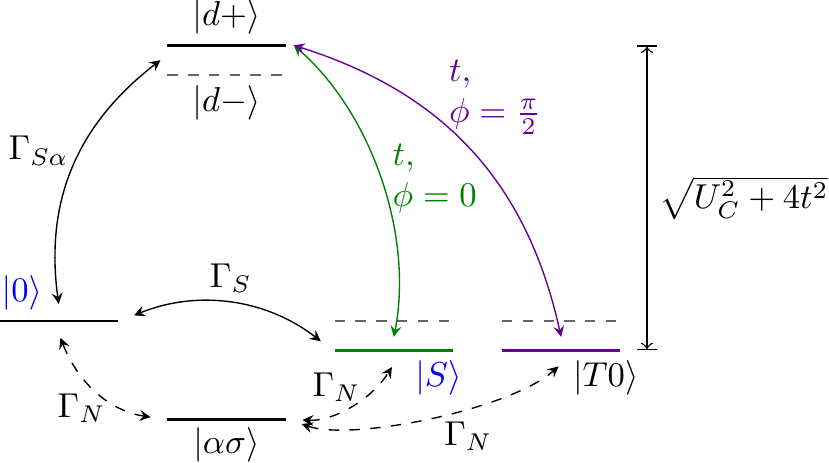}\caption{\label{fig.:eff_level_struc}%
		Effective level structure at the CAR resonance. Finite interdot tunneling ($\phi=0$)
		leads to a level repulsion between the symmetric state $\ket{d+}=(\ket{dR}+\ket{dL})/\sqrt{2}$ 
		and the singlet state $\ket{S}$, see Eq.~\eqref{eq.:hybridized_levels}. SO interaction ($\phi=\pi/2$) 
		leads instead to a level repulsion between the unpolarized triplet state $\ket{T0}$ and the
		symmetric state $\ket{d+}$. The symmetric state, virtually occupied by local Cooper-pair tunneling, plays the role of a dark state.
		}
	\end{figure}
	These observations can be explained by making use of a reduced Hilbert space which describes well the system in the vicinity of the CAR resonance. This simplified model is sketched in Fig.~\ref{fig.:eff_level_struc}.
	The relevant states for the CAR resonance are the empty state $\ket{0}$, the singlet state $\ket{S}$, and the singly occupied states $\ket{\alpha\sigma}$. 
	The states in the even sector $\ket{0}$ and $\ket{S}$ are connected via the nonlocal term $\Gamma_S$, and they are connected to the 
	singly occupied states $\ket{\alpha\sigma}$ via the tunneling rate to the normal lead, $\Gamma_N$. In the absence of interdot tunneling the CAR resonance linewidth is only determined by the nonlocal 
	term $\Gamma_S$, see section \ref{sec.:transportOverview}. However, for finite intradot Coulomb energy in the presence of local terms $\Gamma_{S\alpha}$ and strong interdot tunneling $t$ (with $\phi=0$), we need to consider also another possibility: when the quantum-dots are in the empty state a Cooper pair can  be virtually 
	transferred by means of the local term $\Gamma_{S\alpha}$  in the doubly occupied state $\ket{d\alpha}$ which is converted to the singlet state via the interdot tunneling. One can 
	see in Eq.~\eqref{eq.:Heven} that the tunneling amplitude $(t/\sqrt{2})\cos(\phi)$  couples the $\ket{d\alpha}$ states with
	the singlet state $\ket{S}$. We will quantitatively show that the interference of this alternative channel with the standard nonlocal process fully determines the observed behavior of the CAR peak. 
	
	When $\Gamma_{S}\ll t$, the
	peak shift can be understood in terms of the level repulsion of the singlet state with the doubly occupied state. We first note that the interdot 
	coupling removes the degeneracy of the double occupancies and yields the states $\ket{d\pm}=(\ket{dR}\pm\ket{dL})/\sqrt{2}$. Only the symmetric state  $\ket{d+}$ is affected by the level repulsion with $\ket{S}$.
	In this model the hybridized states $\ket{\pm}\approx\alpha\ket{S}\pm\beta\ket{d+}$ with $\alpha$, $\beta$ c-numbers have the energies\footnote{Only in the limit $\Gamma_S\to0$ the hybridized states $\ket{\pm}$ can be written as linear combination of \ket{S} and \ket{d+}.}
	\begin{align}
	\frac{\epsilon_\pm}{U_C}=\frac{1+4\epsilon/U_C\pm \sqrt{1 + 4\tau^2}}{2}\: ,
	\label{eq.:hybridized_levels}
	\end{align}
	where $\tau=t/U_C$.
	The position of the CAR resonance is the solution of equation $\epsilon_-(\epsilon_{CAR})=0$, the resonance condition between $\ket{-}$ and the empty state $\ket{0}$. The peak position is  $\epsilon_{\textrm{CAR}}=(\sqrt{1 + 4\tau^2}-1)U_C/4$, 
	which fits well the shifting of the peak position [see solid line in Fig.~\ref{fig.:CARpeak}(d)].
	In the limit 
	$U_C\to\infty$ ($\tau\to 0$) the doubly occupied states are unaccessible, even virtually,  and the transport becomes independent of the interdot tunneling.
	
	Finite interdot tunneling also modifies the linewidth of the CAR peak as can be 
	seen in Fig.~\ref{fig.:CARpeak}(a)-(c) and more clearly in Fig.~\ref{fig.:CARpeak}(e)-(f) where we show the linewidth $w_{\textrm{CAR}}$ of the CAR resonance.  
	For $\Gamma_{S}=\sqrt{\Gamma_{SL}\Gamma_{SR}}$ (black circles)
	the width is roughly proportional to the nonlocal coupling
	while for $\Gamma_{S}=0$ (red triangles) it increases with $\tau$. 
	Intriguingly, for an intermediate value of $\Gamma_{S}$ (blue small circles) and $\tau>0$, the linewidth almost vanishes for a specific value of $\tau$ [see Fig.~\ref{fig.:CARpeak}(e)]. This behavior is not seen for $\tau<0$ [see Fig.~\ref{fig.:CARpeak}(f)].
	
	We can explain these results by making use again of the simplified model shown in Fig.~\ref{fig.:eff_level_struc}. 
	In the absence of the interdot tunneling the linewidth of the CAR peak is only determined by the strength of the coupling between the empty 
	state $\ket{0}$ and the singlet state $\ket{S}$. Essentially, it is given by the off-diagonal matrix element 
	$w_{\textrm{CAR}}\approx2\abs{\braketop{0}{H_S}{S}}=\sqrt{2}\Gamma_S$.
	Any additional process that contributes to that coupling between $\ket{0}$ and $\ket{S}$, also through a virtual high energy state, will affect the linewidth. 
	This correction may be obtained considering  the effective   
	Hamiltonian $H_S=H_0+V$, which represents the model shown in Fig.~\ref{fig.:eff_level_struc}, with 
	$H_0=\sum_{i}E_i\ket{i}\bra{i}-(\Gamma_S/\sqrt{2})(\ket{0}\bra{S}+\ket{S}\bra{0})$ for $i=0,S,d+,d-$ and  the 
	perturbation $V=\big[t\ket{d+}\bra{S}-(\Gamma_{S\alpha}/\sqrt{2})\ket{d+}\bra{0}+\mathrm{H.c.}\big]$. Calculating the off-diagonal matrix element up 
	second order in the perturbation, $\mathcal{O}(V^3)$, yields\cite{Berestetskii1982a}
	\begin{align}
\braketop{0}{H_S}{S}=\braketop{0}{H_0}{S}+ \frac{\braketop{0}{V}{d+}\braketop{d+}{V}{S}}{E^{(0)}_S-E^{(0)}_{d+}}
	\end{align}
	with $E^{(0)}_S-E^{(0)}_{d+}=-U_C$. We find for the linewidth $w_{\textrm{CAR}}\approx\sqrt{2}|\Gamma_S-\Gamma_{S\alpha} \tau|$.
	This estimation of the linewidth is indicated by dotted lines in Fig.~\ref{fig.:CARpeak}(e)-(f). It turns out to be a quite good approximation for $\tau\lll 1$ but it worsens for increasing $\Gamma_S$ (see for example the black case $\Gamma_{S}=\sqrt{\Gamma_{SL}\Gamma_{SR}}$). 
	A better approximation, however, is obtained by the substitution $E^{(0)}_S-E^{(0)}_{d+}\to\epsilon_--\epsilon_+=-U_C\sqrt{1 + 4 \tau^2}$ which includes the energy renormalization effects induced by the level repulsion discussed before. 
	Therefore, the linewidth can be approximated by 
	\begin{align}
w_{\textrm{CAR}}\approx\sqrt{2}\Bigg|\Gamma_S -\frac{\Gamma_{S\alpha} \tau}{\sqrt{1 + 4 \tau^2}}\Bigg|
\label{eq:w_car}
	\end{align}
	which fits (solid lines) well the numerical results  based on the full Hamiltonian, see Fig.~\ref{fig.:CARpeak}(e)-(f). 
	Interestingly Eq. ~\eqref{eq:w_car} explains 
	also why for positive (negative)  sign of $t$ the linewidth can decrease (increase) due to destructive (constructive) interference. 
	This is seen comparing the results for $\Gamma_S=0.2\sqrt{\Gamma_{SL}\Gamma_{SR}}$
	in panel (e) and (f) of Fig.~\ref{fig.:CARpeak}.
	Finally, one intriguing consequence of the virtual-state process involving the state $\ket{d+}$ is that it generates nonlocal entangled electrons even in the absence of a direct nonlocal coupling.

	\begin{figure}[t]
		\includegraphics{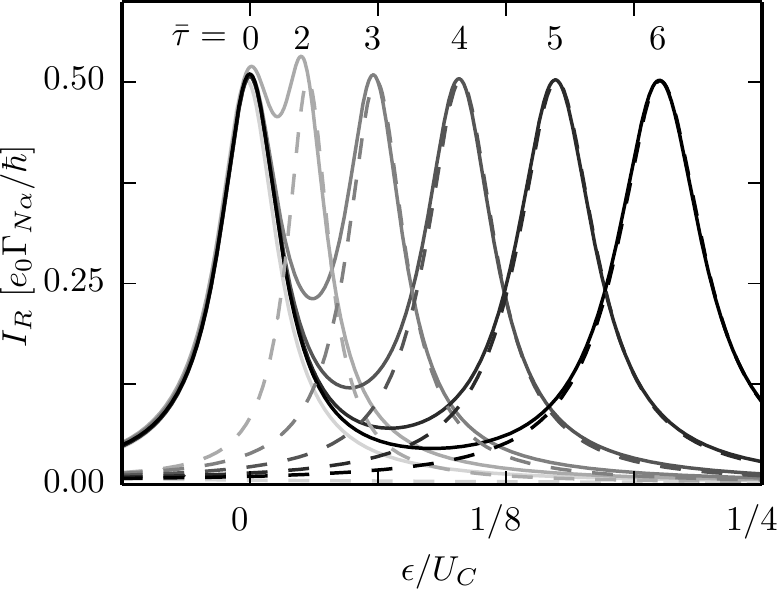}\caption{\label{fig.:so_peak}%
		Current $I_R$ in the CAR resonance as a function of the gate voltage for different values of the tunneling amplitude, $\bar\tau=8\abs{t}/U_C$,  
		$\phi=\pm\pi/2$ for $\Gamma_S=\Gamma_{S\alpha}/3$ (solid line) and $\Gamma_S=0$ (dashed line) keeping fixed the $U_C$. Other parameters as in Fig.~\ref{fig.:CARpeak}.
		}
	\end{figure}
	
	We now turn our attention to the case with the SO coupling and $\phi=\pm \pi/2$.
	We see in panel (b) of Fig.~\ref{fig.:finite_so} that the CAR resonance splits into two lines,  one at $\epsilon\approx0$ and the other shifted towards higher values of $\epsilon$.  In Fig.~\ref{fig.:so_peak} we show the behavior of the CAR peak with increasing values of $t$ for constant values of $U_C$ 
	and $\Gamma_S$.  First we notice that for $\Gamma_S=0$ (dashed lines) the CAR peak does not split but it shifts to the right for increasing values of $\tau$ and no resonance is present at 
	$\epsilon\approx0$.  Instead, for $\Gamma_S\neq 0$, the resonance splits into two resonances, one fixed at $\epsilon\approx0$ and the other right-shifted with $\delta\epsilon_{\textrm{rs}}/U_C\approx (1/2)(t/U_C)^2$.
	This demonstrates the connection with the nonlocal term $\Gamma_S$ of the  CAR peak at $\epsilon\approx0$. 
	
	We numerically observed that the current of the right-shifted peak follows the population of the unpolarized triplet state, $\hbar I_R^{\textrm{rs}}/e_0\Gamma_{NR}\approx2P_{T0}$. These observations suggest that a resonant mechanism involving the virtual occupation of the $\ket{d+}$  is established with the unpolarized triplet state $\ket{T0}$, as depicted  schematically in Fig.~\ref{fig.:eff_level_struc}. This mechanism is analogous to the one induced by the nonlocal singlet proximity in the case of interdot coupling where $\phi=0$. 
	We refer to this resonance as triplet CAR resonance, since it generates  nonlocal entanglement with triplet symmetry. 
	The position and linewidth of this right-shifted resonance  are described by Eq.~\eqref{eq.:hybridized_levels} and Eq.~\eqref{eq:w_car} setting $\Gamma_S=0$, respectively. 
	This is a consequence of the fact that a $s$-wave superconductor cannot induce directly triplet correlations. This shows how the presence of SO coupling can nevertheless induce nonlocal triplet superconducting correlations even when the only superconducting lead has $s$-wave pairing symmetry.\cite{ShekhterPRL2016a,YuPRB2016a}

	\section{\label{sec.:conclusions}Conclusions}
	
	We have presented a comprehensive study of a Cooper-pair splitter based on a double-quantum dot. Employing a master-equation description, in the framework of FCS, 
	we have calculated the current injected into the normal leads. We have considered a finite intra-dot interaction which allows the local transfer of Cooper-pairs from the superconductor to an individual quantum dot. We have studied the signatures of local and nonlocal Andreev reflection in the current injected in the normal leads. The interdot Coulomb interaction separates the local and nonlocal resonances. The effect of interdot tunneling both with and without SO coupling has been considered, too. In particular, we find that the interdot tunneling can induce nonlocal entanglement starting from local Andreev reflection. Furthermore, a process including the virtual doubly-occupied states of the individual dots leads to modifications of the position and linewidth of the current resonances. For the case with SO coupling, we find that a nonlocal triplet pair amplitude can be generated in the system. This mechanism involving the virtual occupation of the doubly occupied states is active only for finite intradot Coulomb interaction.

	\begin{acknowledgments}
		We thank F. Giazotto, S. Roddaro, and S. Kohler for valuable discussions. This work has been supported by Italian's MIUR-FIRB 2012
		via the HybridNanoDev project under Grant no. RBFR1236VV and the EU FP7/2007-2013 under the REA grant agreement no. 630925-COHEAT. A.B. 
		acknowledges support from STM 2015, CNR, the Victoria University of Wellington and the Nano-CNR in Pisa where the work was partially done.
	\end{acknowledgments}
	
	\appendix
	\section{\label{sec.:HS}Matrix representation of the system Hamiltonian}
	In this section we provide the decomposition of the system Hamiltonian $H_S=H_S^{\textrm{even}}\oplus H_S^{\textrm{odd}}$ 
	into sectors with even and odd parity. We assume the single particle level spacing in the quantum dot to be large 
	compared to $U$, $U_\alpha$ and the interdot tunneling $t$, so the total dimension of the system Hilbert space reduces to $16$ states ($8$ even $+$ $8$ odd). 
	Here, we express Eq.~\eqref{eq.:Heff} in the even sector basis $\{ 
	\ket{0}, \ket{S}, \ket{dL}, \ket{dR}, \ket{dd}, 
	\ket{T0}, \ket{T\!\!\uparrow}, \ket{T\!\!\downarrow}
	\}$ stated in table~\ref{tab.:HS_basis}.
	The Hamiltonian for the even charge sector reads
	
	\begin{widetext}
		\begin{align} 
&H_S^{\textrm{even}}=\nonumber\\&{\small
			\begin{pmatrix}
0 & -\frac{1}{\sqrt{2}}\Gamma_{S} & -\frac{1}{2}\Gamma_{SL} & -\frac{1}{2}\Gamma_{SR} & 0 & 0 & 0 & 0\\
-\frac{1}{\sqrt{2}}\Gamma_{S} &\epsilon_L+\epsilon_R+U & \frac{t}{\sqrt{2}}\cos(\phi) & \frac{t}{\sqrt{2}}\cos(\phi) & +\frac{1}{\sqrt{2}}\Gamma_{S} & 0 & 0 & 0\\
-\frac{1}{2}\Gamma_{SL} & \frac{t}{\sqrt{2}}\cos(\phi) & 2 \epsilon_L+U_L & 0 & -\frac{1}{2}\Gamma_{SR} & i\frac{t}{\sqrt{2}}\sin(\phi)  & 0 & 0\\
-\frac{1}{2}\Gamma_{SR} & \frac{t}{\sqrt{2}}\cos(\phi) & 0 & 2 \epsilon_R+U_R & -\frac{1}{2}\Gamma_{SL} & i\frac{t}{\sqrt{2}}\sin(\phi) & 0 & 0\\
0 & +\frac{1}{\sqrt{2}}\Gamma_{S} & -\frac{1}{2}\Gamma_{SR} & -\frac{1}{2}\Gamma_{SL} & 2(\epsilon_R+\epsilon_L)+U_R+U_L+4U & 0 & 0 & 0\\
0 & 0 & -i\frac{t}{\sqrt{2}}\sin(\phi) & -i\frac{t}{\sqrt{2}}\sin(\phi) & 0 & \epsilon_L+\epsilon_R+U  & 0 & 0\\
0 & 0 & 0 & 0 & 0 & 0 & \epsilon_L+\epsilon_R+U  & 0\\
0 & 0 & 0 & 0 & 0 & 0 & 0 & \epsilon_L+\epsilon_R+U  \\ 
			\end{pmatrix}}.
\label{eq.:Heven}
		\end{align}%
		Note that the interdot 
		tunneling preserves the spin of the tunneling electrons (being time reversal invariant) and the total parity of the DQD.
		In absence of the SO interaction, $\phi=0$, all triplet states $\ket{Ti}$ are completely decoupled from the other even parity states. When $\phi\neq\pi k$, where $k$ integer, the unpolarized triplet state $\ket{T0}$ couples with the doubly occupied states $\ket{d\alpha}$.
		The Hamiltonian for the odd charge sector, in the basis
		$\{
		\ket{R\!\!\uparrow}, \ket{R\!\!\downarrow}, \ket{L\!\!\uparrow}, \ket{L\!\!\downarrow},  
		\ket{tR\!\!\uparrow}, \ket{tR\!\!\downarrow}, \ket{tL\!\!\uparrow},\\ \ket{tL\!\!\downarrow}
		\}$, is given by
		\begin{align}
			H_S^{\text{odd}}=\begin{pmatrix}
\epsilon_R & 0 & \frac{t}{2} e^{-i\phi} & 0 & -\frac{1}{2}\Gamma_{SL} & 0 & +\frac{1}{2}\Gamma_{S} & 0\\
0 &\epsilon_R & 0 & \frac{t}{2} e^{+i\phi} & 0 & -\frac{1}{2}\Gamma_{SL} & 0 & +\frac{1}{2}\Gamma_{S} \\
\frac{t}{2} e^{+i\phi} & 0 & \epsilon_L & 0 &  +\frac{1}{2}\Gamma_{S} & 0 & -\frac{1}{2}\Gamma_{SR} & 0\\
0 & \frac{t}{2} e^{-i\phi} & 0 & \epsilon_L & 0 & +\frac{1}{2}\Gamma_{S} & 0 &-\frac{1}{2}\Gamma_{SR}\\
-\frac{1}{2}\Gamma_{SL} & 0  & +\frac{1}{2}\Gamma_{S}  & 0 & E_{tR\uparrow} & 0  & -\frac{t}{2} e^{-i\phi}  & 0\\
0 & -\frac{1}{2}\Gamma_{SL}   & 0 & +\frac{1}{2}\Gamma_{S} & 0 & E_{tR\downarrow} & 0  & -\frac{t}{2} e^{+i\phi}\\
+\frac{1}{2}\Gamma_{S} & 0  & -\frac{1}{2}\Gamma_{SR} & 0 & -\frac{t}{2} e^{+i\phi} & 0 & E_{tL\uparrow}  & 0\\
0 & +\frac{1}{2}\Gamma_{S} & 0  & -\frac{1}{2}\Gamma_{SR} & 0 & -\frac{t}{2} e^{-i\phi} & 0 & E_{tL\downarrow} \\
			\end{pmatrix},
\label{eq.:Hodd}
		\end{align}%
		where $E_{t\alpha\sigma}=2\epsilon_{\bar{\alpha}} +U_{\bar{\alpha}}+\epsilon_\alpha+2U$ with 
		$\alpha=R,L$ ($\bar{\alpha}=L,R$) and $\sigma=\uparrow,\downarrow$. 
	\end{widetext}


\begin{thebibliography}{70}%
	\makeatletter
	\providecommand \@ifxundefined [1]{%
	 \@ifx{#1\undefined}
	}%
	\providecommand \@ifnum [1]{%
	 \ifnum #1\expandafter \@firstoftwo
	 \else \expandafter \@secondoftwo
	 \fi
	}%
	\providecommand \@ifx [1]{%
	 \ifx #1\expandafter \@firstoftwo
	 \else \expandafter \@secondoftwo
	 \fi
	}%
	\providecommand \natexlab [1]{#1}%
	\providecommand \enquote  [1]{``#1''}%
	\providecommand \bibnamefont  [1]{#1}%
	\providecommand \bibfnamefont [1]{#1}%
	\providecommand \citenamefont [1]{#1}%
	\providecommand \href@noop [0]{\@secondoftwo}%
	\providecommand \href [0]{\begingroup \@sanitize@url \@href}%
	\providecommand \@href[1]{\@@startlink{#1}\@@href}%
	\providecommand \@@href[1]{\endgroup#1\@@endlink}%
	\providecommand \@sanitize@url [0]{\catcode `\\12\catcode `\$12\catcode
	  `\&12\catcode `\#12\catcode `\^12\catcode `\_12\catcode `\%12\relax}%
	\providecommand \@@startlink[1]{}%
	\providecommand \@@endlink[0]{}%
	\providecommand \url  [0]{\begingroup\@sanitize@url \@url }%
	\providecommand \@url [1]{\endgroup\@href {#1}{\urlprefix }}%
	\providecommand \urlprefix  [0]{URL }%
	\providecommand \Eprint [0]{\href }%
	\providecommand \doibase [0]{http://dx.doi.org/}%
	\providecommand \selectlanguage [0]{\@gobble}%
	\providecommand \bibinfo  [0]{\@secondoftwo}%
	\providecommand \bibfield  [0]{\@secondoftwo}%
	\providecommand \translation [1]{[#1]}%
	\providecommand \BibitemOpen [0]{}%
	\providecommand \bibitemStop [0]{}%
	\providecommand \bibitemNoStop [0]{.\EOS\space}%
	\providecommand \EOS [0]{\spacefactor3000\relax}%
	\providecommand \BibitemShut  [1]{\csname bibitem#1\endcsname}%
	\let\auto@bib@innerbib\@empty
	\bibitem [{\citenamefont {Monroe}(2002)}]{MonroeNature2002a}%
	  \BibitemOpen
	  \bibfield  {author} {\bibinfo {author} {\bibfnamefont {C.}~\bibnamefont
	  {Monroe}},\ }\href@noop {} {\bibfield  {journal} {\bibinfo  {journal}
	  {Nature}\ }\textbf {\bibinfo {volume} {416}},\ \bibinfo {pages} {238}
	  (\bibinfo {year} {2002})}\BibitemShut {NoStop}%
	\bibitem [{\citenamefont {Martinis}(2009)}]{MartinisQuantumIP2009a}%
	  \BibitemOpen
	  \bibfield  {author} {\bibinfo {author} {\bibfnamefont {J.}~\bibnamefont
	  {Martinis}},\ }\href@noop {} {\bibfield  {journal} {\bibinfo  {journal}
	  {Quantum Inf. Process.}\ }\textbf {\bibinfo {volume} {8}},\ \bibinfo {pages}
	  {81} (\bibinfo {year} {2009})}\BibitemShut {NoStop}%
	\bibitem [{\citenamefont {Ladd}\ \emph {et~al.}(2010)\citenamefont {Ladd},
	  \citenamefont {Jelezko}, \citenamefont {Laflamme}, \citenamefont {Nakamura},
	  \citenamefont {Monroe},\ and\ \citenamefont {O'Brien}}]{LaddNature2010a}%
	  \BibitemOpen
	  \bibfield  {author} {\bibinfo {author} {\bibfnamefont {T.~D.}\ \bibnamefont
	  {Ladd}}, \bibinfo {author} {\bibfnamefont {F.}~\bibnamefont {Jelezko}},
	  \bibinfo {author} {\bibfnamefont {R.}~\bibnamefont {Laflamme}}, \bibinfo
	  {author} {\bibfnamefont {Y.}~\bibnamefont {Nakamura}}, \bibinfo {author}
	  {\bibfnamefont {C.}~\bibnamefont {Monroe}}, \ and\ \bibinfo {author}
	  {\bibfnamefont {J.~L.}\ \bibnamefont {O'Brien}},\ }\href@noop {} {\bibfield
	  {journal} {\bibinfo  {journal} {Nature}\ }\textbf {\bibinfo {volume} {464}},\
	  \bibinfo {pages} {45} (\bibinfo {year} {2010})}\BibitemShut {NoStop}%
	\bibitem [{\citenamefont {Lo}\ \emph {et~al.}(2014)\citenamefont {Lo},
	  \citenamefont {Curty},\ and\ \citenamefont {Tamaki}}]{LoNatureP2014a}%
	  \BibitemOpen
	  \bibfield  {author} {\bibinfo {author} {\bibfnamefont {H.-K.}\ \bibnamefont
	  {Lo}}, \bibinfo {author} {\bibfnamefont {M.}~\bibnamefont {Curty}}, \ and\
	  \bibinfo {author} {\bibfnamefont {K.}~\bibnamefont {Tamaki}},\ }\href@noop {}
	  {\bibfield  {journal} {\bibinfo  {journal} {Nature Photon.}\ }\textbf
	  {\bibinfo {volume} {8}},\ \bibinfo {pages} {595} (\bibinfo {year}
	  {2014})}\BibitemShut {NoStop}%
	\bibitem [{\citenamefont {Linder}\ and\ \citenamefont
	  {Robinson}(2015)}]{LinderNatPhys2015a}%
	  \BibitemOpen
	  \bibfield  {author} {\bibinfo {author} {\bibfnamefont {J.}~\bibnamefont
	  {Linder}}\ and\ \bibinfo {author} {\bibfnamefont {J.~W.~A.}\ \bibnamefont
	  {Robinson}},\ }\href@noop {} {\bibfield  {journal} {\bibinfo  {journal}
	  {Nature Phys.}\ }\textbf {\bibinfo {volume} {11}},\ \bibinfo {pages} {307}
	  (\bibinfo {year} {2015})}\BibitemShut {NoStop}%
	\bibitem [{\citenamefont {Yoshimi}\ \emph {et~al.}(2015)\citenamefont
	  {Yoshimi}, \citenamefont {Tsukazaki}, \citenamefont {Kozuka}, \citenamefont
	  {Falson}, \citenamefont {Takahashi}, \citenamefont {Checkelsky},
	  \citenamefont {Nagaosa}, \citenamefont {Kawasaki},\ and\ \citenamefont
	  {Tokura}}]{YoshimiNatureC2015a}%
	  \BibitemOpen
	  \bibfield  {author} {\bibinfo {author} {\bibfnamefont {R.}~\bibnamefont
	  {Yoshimi}}, \bibinfo {author} {\bibfnamefont {A.}~\bibnamefont {Tsukazaki}},
	  \bibinfo {author} {\bibfnamefont {Y.}~\bibnamefont {Kozuka}}, \bibinfo
	  {author} {\bibfnamefont {J.}~\bibnamefont {Falson}}, \bibinfo {author}
	  {\bibfnamefont {K.}~\bibnamefont {Takahashi}}, \bibinfo {author}
	  {\bibfnamefont {J.}~\bibnamefont {Checkelsky}}, \bibinfo {author}
	  {\bibfnamefont {N.}~\bibnamefont {Nagaosa}}, \bibinfo {author} {\bibfnamefont
	  {M.}~\bibnamefont {Kawasaki}}, \ and\ \bibinfo {author} {\bibfnamefont
	  {Y.}~\bibnamefont {Tokura}},\ }\href@noop {} {\bibfield  {journal} {\bibinfo
	  {journal} {Nature Commun.}\ }\textbf {\bibinfo {volume} {6}},\ \bibinfo
	  {pages} {6627} (\bibinfo {year} {2015})}\BibitemShut {NoStop}%
	\bibitem [{\citenamefont {Qi}\ and\ \citenamefont {Zhang}(2011)}]{QiRMP2011a}%
	  \BibitemOpen
	  \bibfield  {author} {\bibinfo {author} {\bibfnamefont {X.-L.}\ \bibnamefont
	  {Qi}}\ and\ \bibinfo {author} {\bibfnamefont {S.-C.}\ \bibnamefont {Zhang}},\
	  }\href@noop {} {\bibfield  {journal} {\bibinfo  {journal} {Rev. Mod. Phys.}\
	  }\textbf {\bibinfo {volume} {83}},\ \bibinfo {pages} {1057} (\bibinfo {year}
	  {2011})}\BibitemShut {NoStop}%
	\bibitem [{\citenamefont {Kitaev}(2001)}]{KitaevPU2001a}%
	  \BibitemOpen
	  \bibfield  {author} {\bibinfo {author} {\bibfnamefont {A.}~\bibnamefont
	  {Kitaev}},\ }\href@noop {} {\bibfield  {journal} {\bibinfo  {journal} {Phys.
	  Usp.}\ }\textbf {\bibinfo {volume} {44}},\ \bibinfo {pages} {131} (\bibinfo
	  {year} {2001})}\BibitemShut {NoStop}%
	\bibitem [{\citenamefont {Nayak}\ \emph {et~al.}(2008)\citenamefont {Nayak},
	  \citenamefont {Simon}, \citenamefont {Stern}, \citenamefont {Freedman},\ and\
	  \citenamefont {Das~Sarma}}]{NayakRMP2008a}%
	  \BibitemOpen
	  \bibfield  {author} {\bibinfo {author} {\bibfnamefont {C.}~\bibnamefont
	  {Nayak}}, \bibinfo {author} {\bibfnamefont {S.~H.}\ \bibnamefont {Simon}},
	  \bibinfo {author} {\bibfnamefont {A.}~\bibnamefont {Stern}}, \bibinfo
	  {author} {\bibfnamefont {M.}~\bibnamefont {Freedman}}, \ and\ \bibinfo
	  {author} {\bibfnamefont {S.}~\bibnamefont {Das~Sarma}},\ }\href@noop {}
	  {\bibfield  {journal} {\bibinfo  {journal} {Rev. Mod. Phys.}\ }\textbf
	  {\bibinfo {volume} {80}},\ \bibinfo {pages} {1083} (\bibinfo {year}
	  {2008})}\BibitemShut {NoStop}%
	\bibitem [{\citenamefont {Alicea}(2012)}]{AliceRPP2012a}%
	  \BibitemOpen
	  \bibfield  {author} {\bibinfo {author} {\bibfnamefont {J.}~\bibnamefont
	  {Alicea}},\ }\href@noop {} {\bibfield  {journal} {\bibinfo  {journal} {Rep.
	  Prog. Phys.}\ }\textbf {\bibinfo {volume} {75}},\ \bibinfo {pages} {076501}
	  (\bibinfo {year} {2012})}\BibitemShut {NoStop}%
	\bibitem [{\citenamefont {Romeo}\ \emph {et~al.}(2012)\citenamefont {Romeo},
	  \citenamefont {Roddaro}, \citenamefont {Pitanti}, \citenamefont {Ercolani},
	  \citenamefont {Sorba},\ and\ \citenamefont {Beltram}}]{RomeoNanoL2012a}%
	  \BibitemOpen
	  \bibfield  {author} {\bibinfo {author} {\bibfnamefont {L.}~\bibnamefont
	  {Romeo}}, \bibinfo {author} {\bibfnamefont {S.}~\bibnamefont {Roddaro}},
	  \bibinfo {author} {\bibfnamefont {A.}~\bibnamefont {Pitanti}}, \bibinfo
	  {author} {\bibfnamefont {D.}~\bibnamefont {Ercolani}}, \bibinfo {author}
	  {\bibfnamefont {L.}~\bibnamefont {Sorba}}, \ and\ \bibinfo {author}
	  {\bibfnamefont {F.}~\bibnamefont {Beltram}},\ }\href@noop {} {\bibfield
	  {journal} {\bibinfo  {journal} {Nano Lett.}\ }\textbf {\bibinfo {volume}
	  {12}},\ \bibinfo {pages} {4490} (\bibinfo {year} {2012})}\BibitemShut
	  {NoStop}%
	\bibitem [{\citenamefont {Rossella}\ \emph {et~al.}(2014)\citenamefont
	  {Rossella}, \citenamefont {Bertoni}, \citenamefont {Ercolani}, \citenamefont
	  {Rontani}, \citenamefont {Sorba}, \citenamefont {Beltram},\ and\
	  \citenamefont {Roddaro}}]{RossellaNatNano2014a}%
	  \BibitemOpen
	  \bibfield  {author} {\bibinfo {author} {\bibfnamefont {F.}~\bibnamefont
	  {Rossella}}, \bibinfo {author} {\bibfnamefont {A.}~\bibnamefont {Bertoni}},
	  \bibinfo {author} {\bibfnamefont {D.}~\bibnamefont {Ercolani}}, \bibinfo
	  {author} {\bibfnamefont {M.}~\bibnamefont {Rontani}}, \bibinfo {author}
	  {\bibfnamefont {L.}~\bibnamefont {Sorba}}, \bibinfo {author} {\bibfnamefont
	  {F.}~\bibnamefont {Beltram}}, \ and\ \bibinfo {author} {\bibfnamefont
	  {S.}~\bibnamefont {Roddaro}},\ }\href@noop {} {\bibfield  {journal} {\bibinfo
	   {journal} {Nature Nano}\ }\textbf {\bibinfo {volume} {9}},\ \bibinfo {pages}
	  {997} (\bibinfo {year} {2014})}\BibitemShut {NoStop}%
	\bibitem [{\citenamefont {Giazotto}\ \emph {et~al.}(2011)\citenamefont
	  {Giazotto}, \citenamefont {Spathis}, \citenamefont {Roddaro}, \citenamefont
	  {Biswas}, \citenamefont {Taddei}, \citenamefont {Governale},\ and\
	  \citenamefont {Sorba}}]{GiazottoNatPhys2011a}%
	  \BibitemOpen
	  \bibfield  {author} {\bibinfo {author} {\bibfnamefont {F.}~\bibnamefont
	  {Giazotto}}, \bibinfo {author} {\bibfnamefont {P.}~\bibnamefont {Spathis}},
	  \bibinfo {author} {\bibfnamefont {S.}~\bibnamefont {Roddaro}}, \bibinfo
	  {author} {\bibfnamefont {S.}~\bibnamefont {Biswas}}, \bibinfo {author}
	  {\bibfnamefont {F.}~\bibnamefont {Taddei}}, \bibinfo {author} {\bibfnamefont
	  {M.}~\bibnamefont {Governale}}, \ and\ \bibinfo {author} {\bibfnamefont
	  {L.}~\bibnamefont {Sorba}},\ }\href@noop {} {\bibfield  {journal} {\bibinfo
	  {journal} {Nature. Phys.}\ }\textbf {\bibinfo {volume} {7}},\ \bibinfo
	  {pages} {857} (\bibinfo {year} {2011})}\BibitemShut {NoStop}%
	\bibitem [{\citenamefont {Roddaro}\ \emph {et~al.}(2011)\citenamefont
	  {Roddaro}, \citenamefont {Pescaglini}, \citenamefont {Ercolani},
	  \citenamefont {Sorba}, \citenamefont {Giazotto},\ and\ \citenamefont
	  {Beltram}}]{RoddaroNanoR2011a}%
	  \BibitemOpen
	  \bibfield  {author} {\bibinfo {author} {\bibfnamefont {S.}~\bibnamefont
	  {Roddaro}}, \bibinfo {author} {\bibfnamefont {A.}~\bibnamefont {Pescaglini}},
	  \bibinfo {author} {\bibfnamefont {D.}~\bibnamefont {Ercolani}}, \bibinfo
	  {author} {\bibfnamefont {L.}~\bibnamefont {Sorba}}, \bibinfo {author}
	  {\bibfnamefont {F.}~\bibnamefont {Giazotto}}, \ and\ \bibinfo {author}
	  {\bibfnamefont {F.}~\bibnamefont {Beltram}},\ }\href@noop {} {\bibfield
	  {journal} {\bibinfo  {journal} {Nano Res.}\ }\textbf {\bibinfo {volume}
	  {4}},\ \bibinfo {pages} {259} (\bibinfo {year} {2011})}\BibitemShut {NoStop}%
	\bibitem [{\citenamefont {Spathis}\ \emph {et~al.}(2011)\citenamefont
	  {Spathis}, \citenamefont {Biswas}, \citenamefont {Roddaro}, \citenamefont
	  {Sorba}, \citenamefont {Giazotto},\ and\ \citenamefont
	  {Beltram}}]{SpathisNanotechnology2011a}%
	  \BibitemOpen
	  \bibfield  {author} {\bibinfo {author} {\bibfnamefont {P.}~\bibnamefont
	  {Spathis}}, \bibinfo {author} {\bibfnamefont {S.}~\bibnamefont {Biswas}},
	  \bibinfo {author} {\bibfnamefont {S.}~\bibnamefont {Roddaro}}, \bibinfo
	  {author} {\bibfnamefont {L.}~\bibnamefont {Sorba}}, \bibinfo {author}
	  {\bibfnamefont {F.}~\bibnamefont {Giazotto}}, \ and\ \bibinfo {author}
	  {\bibfnamefont {F.}~\bibnamefont {Beltram}},\ }\href@noop {} {\bibfield
	  {journal} {\bibinfo  {journal} {Nanotechnology}\ }\textbf {\bibinfo {volume}
	  {22}},\ \bibinfo {pages} {105201} (\bibinfo {year} {2011})}\BibitemShut
	  {NoStop}%
	\bibitem [{\citenamefont {Lee}\ \emph {et~al.}(2014)\citenamefont {Lee},
	  \citenamefont {Jiang}, \citenamefont {Houzet}, \citenamefont {Aguado},
	  \citenamefont {Lieber},\ and\ \citenamefont
	  {De~Franceschi}}]{LeeNatNano2014a}%
	  \BibitemOpen
	  \bibfield  {author} {\bibinfo {author} {\bibfnamefont {E.~J.~H.}\
	  \bibnamefont {Lee}}, \bibinfo {author} {\bibfnamefont {X.}~\bibnamefont
	  {Jiang}}, \bibinfo {author} {\bibfnamefont {M.}~\bibnamefont {Houzet}},
	  \bibinfo {author} {\bibfnamefont {R.}~\bibnamefont {Aguado}}, \bibinfo
	  {author} {\bibfnamefont {C.~M.}\ \bibnamefont {Lieber}}, \ and\ \bibinfo
	  {author} {\bibfnamefont {S.}~\bibnamefont {De~Franceschi}},\ }\href@noop {}
	  {\bibfield  {journal} {\bibinfo  {journal} {Nature Nano}\ }\textbf {\bibinfo
	  {volume} {9}},\ \bibinfo {pages} {79} (\bibinfo {year} {2014})}\BibitemShut
	  {NoStop}%
	\bibitem [{\citenamefont {Yokoyama}\ \emph {et~al.}(2014)\citenamefont
	  {Yokoyama}, \citenamefont {Eto},\ and\ \citenamefont
	  {Nazarov}}]{YokoyamaPRB2014a}%
	  \BibitemOpen
	  \bibfield  {author} {\bibinfo {author} {\bibfnamefont {T.}~\bibnamefont
	  {Yokoyama}}, \bibinfo {author} {\bibfnamefont {M.}~\bibnamefont {Eto}}, \
	  and\ \bibinfo {author} {\bibfnamefont {Y.~V.}\ \bibnamefont {Nazarov}},\
	  }\href@noop {} {\bibfield  {journal} {\bibinfo  {journal} {Phys. Rev. B}\
	  }\textbf {\bibinfo {volume} {89}},\ \bibinfo {pages} {195407} (\bibinfo
	  {year} {2014})}\BibitemShut {NoStop}%
	\bibitem [{\citenamefont {Mironov}\ \emph {et~al.}(2015)\citenamefont
	  {Mironov}, \citenamefont {Mel'nikov},\ and\ \citenamefont
	  {Buzdin}}]{MironovPRL2015a}%
	  \BibitemOpen
	  \bibfield  {author} {\bibinfo {author} {\bibfnamefont {S.~V.}\ \bibnamefont
	  {Mironov}}, \bibinfo {author} {\bibfnamefont {A.~S.}\ \bibnamefont
	  {Mel'nikov}}, \ and\ \bibinfo {author} {\bibfnamefont {A.~I.}\ \bibnamefont
	  {Buzdin}},\ }\href@noop {} {\bibfield  {journal} {\bibinfo  {journal} {Phys.
	  Rev. Lett.}\ }\textbf {\bibinfo {volume} {114}},\ \bibinfo {pages} {227001}
	  (\bibinfo {year} {2015})}\BibitemShut {NoStop}%
	\bibitem [{\citenamefont {{\v{Z}}onda}\ \emph {et~al.}(2015)\citenamefont
	  {{\v{Z}}onda}, \citenamefont {Pokorn{\`y}}, \citenamefont {Jani{\v{s}}},\
	  and\ \citenamefont {Novotn{\`y}}}]{ZondaSR2015a}%
	  \BibitemOpen
	  \bibfield  {author} {\bibinfo {author} {\bibfnamefont {M.}~\bibnamefont
	  {{\v{Z}}onda}}, \bibinfo {author} {\bibfnamefont {V.}~\bibnamefont
	  {Pokorn{\`y}}}, \bibinfo {author} {\bibfnamefont {V.}~\bibnamefont
	  {Jani{\v{s}}}}, \ and\ \bibinfo {author} {\bibfnamefont {T.}~\bibnamefont
	  {Novotn{\`y}}},\ }\href@noop {} {\bibfield  {journal} {\bibinfo  {journal}
	  {Sci. Rep.}\ }\textbf {\bibinfo {volume} {5}},\ \bibinfo {pages} {8821}
	  (\bibinfo {year} {2015})}\BibitemShut {NoStop}%
	\bibitem [{\citenamefont {Marra}\ \emph {et~al.}(2016)\citenamefont {Marra},
	  \citenamefont {Citro},\ and\ \citenamefont {Braggio}}]{MarraPRB2016a}%
	  \BibitemOpen
	  \bibfield  {author} {\bibinfo {author} {\bibfnamefont {P.}~\bibnamefont
	  {Marra}}, \bibinfo {author} {\bibfnamefont {R.}~\bibnamefont {Citro}}, \ and\
	  \bibinfo {author} {\bibfnamefont {A.}~\bibnamefont {Braggio}},\ }\href@noop
	  {} {\bibfield  {journal} {\bibinfo  {journal} {Phys. Rev. B}\ }\textbf
	  {\bibinfo {volume} {93}},\ \bibinfo {pages} {220507} (\bibinfo {year}
	  {2016})}\BibitemShut {NoStop}%
	\bibitem [{\citenamefont {Shekhter}\ \emph {et~al.}(2016)\citenamefont
	  {Shekhter}, \citenamefont {Entin-Wohlman}, \citenamefont {Jonson},\ and\
	  \citenamefont {Aharony}}]{ShekhterPRL2016a}%
	  \BibitemOpen
	  \bibfield  {author} {\bibinfo {author} {\bibfnamefont {R.~I.}\ \bibnamefont
	  {Shekhter}}, \bibinfo {author} {\bibfnamefont {O.}~\bibnamefont
	  {Entin-Wohlman}}, \bibinfo {author} {\bibfnamefont {M.}~\bibnamefont
	  {Jonson}}, \ and\ \bibinfo {author} {\bibfnamefont {A.}~\bibnamefont
	  {Aharony}},\ }\href@noop {} {\bibfield  {journal} {\bibinfo  {journal} {Phys.
	  Rev. Lett.}\ }\textbf {\bibinfo {volume} {116}},\ \bibinfo {pages} {217001}
	  (\bibinfo {year} {2016})}\BibitemShut {NoStop}%
	\bibitem [{\citenamefont {Yu}\ and\ \citenamefont {Wu}(2016)}]{YuPRB2016a}%
	  \BibitemOpen
	  \bibfield  {author} {\bibinfo {author} {\bibfnamefont {T.}~\bibnamefont
	  {Yu}}\ and\ \bibinfo {author} {\bibfnamefont {M.~W.}\ \bibnamefont {Wu}},\
	  }\href@noop {} {\bibfield  {journal} {\bibinfo  {journal} {Phys. Rev. B}\
	  }\textbf {\bibinfo {volume} {93}},\ \bibinfo {pages} {195308} (\bibinfo
	  {year} {2016})}\BibitemShut {NoStop}%
	\bibitem [{\citenamefont {Bouchiat}\ \emph {et~al.}(2003)\citenamefont
	  {Bouchiat}, \citenamefont {Chtchelkatchev}, \citenamefont {Feinberg},
	  \citenamefont {Lesovik}, \citenamefont {Martin},\ and\ \citenamefont
	  {Torr\`es}}]{BouchiatNanotechnology2003a}%
	  \BibitemOpen
	  \bibfield  {author} {\bibinfo {author} {\bibfnamefont {V.}~\bibnamefont
	  {Bouchiat}}, \bibinfo {author} {\bibfnamefont {N.}~\bibnamefont
	  {Chtchelkatchev}}, \bibinfo {author} {\bibfnamefont {D.}~\bibnamefont
	  {Feinberg}}, \bibinfo {author} {\bibfnamefont {G.~B.}\ \bibnamefont
	  {Lesovik}}, \bibinfo {author} {\bibfnamefont {T.}~\bibnamefont {Martin}}, \
	  and\ \bibinfo {author} {\bibfnamefont {J.}~\bibnamefont {Torr\`es}},\
	  }\href@noop {} {\bibfield  {journal} {\bibinfo  {journal} {Nanotechnology}\
	  }\textbf {\bibinfo {volume} {14}},\ \bibinfo {pages} {77} (\bibinfo {year}
	  {2003})}\BibitemShut {NoStop}%
	\bibitem [{\citenamefont {Russo}\ \emph {et~al.}(2005)\citenamefont {Russo},
	  \citenamefont {Kroug}, \citenamefont {Klapwijk},\ and\ \citenamefont
	  {Morpurgo}}]{RussoPRL2005a}%
	  \BibitemOpen
	  \bibfield  {author} {\bibinfo {author} {\bibfnamefont {S.}~\bibnamefont
	  {Russo}}, \bibinfo {author} {\bibfnamefont {M.}~\bibnamefont {Kroug}},
	  \bibinfo {author} {\bibfnamefont {T.~M.}\ \bibnamefont {Klapwijk}}, \ and\
	  \bibinfo {author} {\bibfnamefont {A.~F.}\ \bibnamefont {Morpurgo}},\
	  }\href@noop {} {\bibfield  {journal} {\bibinfo  {journal} {Phys. Rev. Lett.}\
	  }\textbf {\bibinfo {volume} {95}},\ \bibinfo {pages} {027002} (\bibinfo
	  {year} {2005})}\BibitemShut {NoStop}%
	\bibitem [{\citenamefont {Schindele}\ \emph {et~al.}(2012)\citenamefont
	  {Schindele}, \citenamefont {Baumgartner},\ and\ \citenamefont
	  {Sch\"onenberger}}]{SchindelePRL2012a}%
	  \BibitemOpen
	  \bibfield  {author} {\bibinfo {author} {\bibfnamefont {J.}~\bibnamefont
	  {Schindele}}, \bibinfo {author} {\bibfnamefont {A.}~\bibnamefont
	  {Baumgartner}}, \ and\ \bibinfo {author} {\bibfnamefont {C.}~\bibnamefont
	  {Sch\"onenberger}},\ }\href@noop {} {\bibfield  {journal} {\bibinfo
	  {journal} {Phys. Rev. Lett.}\ }\textbf {\bibinfo {volume} {109}},\ \bibinfo
	  {pages} {157002} (\bibinfo {year} {2012})}\BibitemShut {NoStop}%
	\bibitem [{\citenamefont {F\"ul\"op}\ \emph {et~al.}(2014)\citenamefont
	  {F\"ul\"op}, \citenamefont {d'Hollosy}, \citenamefont {Baumgartner},
	  \citenamefont {Makk}, \citenamefont {Guzenko}, \citenamefont {Madsen},
	  \citenamefont {Nyg\aa{}rd}, \citenamefont {Sch\"onenberger},\ and\
	  \citenamefont {Csonka}}]{FueloepPRB2014a}%
	  \BibitemOpen
	  \bibfield  {author} {\bibinfo {author} {\bibfnamefont {G.}~\bibnamefont
	  {F\"ul\"op}}, \bibinfo {author} {\bibfnamefont {S.}~\bibnamefont
	  {d'Hollosy}}, \bibinfo {author} {\bibfnamefont {A.}~\bibnamefont
	  {Baumgartner}}, \bibinfo {author} {\bibfnamefont {P.}~\bibnamefont {Makk}},
	  \bibinfo {author} {\bibfnamefont {V.~A.}\ \bibnamefont {Guzenko}}, \bibinfo
	  {author} {\bibfnamefont {M.~H.}\ \bibnamefont {Madsen}}, \bibinfo {author}
	  {\bibfnamefont {J.}~\bibnamefont {Nyg\aa{}rd}}, \bibinfo {author}
	  {\bibfnamefont {C.}~\bibnamefont {Sch\"onenberger}}, \ and\ \bibinfo {author}
	  {\bibfnamefont {S.}~\bibnamefont {Csonka}},\ }\href@noop {} {\bibfield
	  {journal} {\bibinfo  {journal} {Phys. Rev. B}\ }\textbf {\bibinfo {volume}
	  {90}},\ \bibinfo {pages} {235412} (\bibinfo {year} {2014})}\BibitemShut
	  {NoStop}%
	\bibitem [{\citenamefont {Das}\ \emph {et~al.}(2012)\citenamefont {Das},
	  \citenamefont {Ronen}, \citenamefont {Heiblum}, \citenamefont {Mahalu},
	  \citenamefont {Kretinin},\ and\ \citenamefont {Shtrikman}}]{DasNatureC2012a}%
	  \BibitemOpen
	  \bibfield  {author} {\bibinfo {author} {\bibfnamefont {A.}~\bibnamefont
	  {Das}}, \bibinfo {author} {\bibfnamefont {Y.}~\bibnamefont {Ronen}}, \bibinfo
	  {author} {\bibfnamefont {M.}~\bibnamefont {Heiblum}}, \bibinfo {author}
	  {\bibfnamefont {D.}~\bibnamefont {Mahalu}}, \bibinfo {author} {\bibfnamefont
	  {A.~V.}\ \bibnamefont {Kretinin}}, \ and\ \bibinfo {author} {\bibfnamefont
	  {H.}~\bibnamefont {Shtrikman}},\ }\href@noop {} {\bibfield  {journal}
	  {\bibinfo  {journal} {Nature Commun.}\ }\textbf {\bibinfo {volume} {3}},\
	  \bibinfo {pages} {1165} (\bibinfo {year} {2012})}\BibitemShut {NoStop}%
	\bibitem [{\citenamefont {Schindele}\ \emph {et~al.}(2014)\citenamefont
	  {Schindele}, \citenamefont {Baumgartner}, \citenamefont {Maurand},
	  \citenamefont {Weiss},\ and\ \citenamefont
	  {Sch\"onenberger}}]{SchindelePRB2014a}%
	  \BibitemOpen
	  \bibfield  {author} {\bibinfo {author} {\bibfnamefont {J.}~\bibnamefont
	  {Schindele}}, \bibinfo {author} {\bibfnamefont {A.}~\bibnamefont
	  {Baumgartner}}, \bibinfo {author} {\bibfnamefont {R.}~\bibnamefont
	  {Maurand}}, \bibinfo {author} {\bibfnamefont {M.}~\bibnamefont {Weiss}}, \
	  and\ \bibinfo {author} {\bibfnamefont {C.}~\bibnamefont {Sch\"onenberger}},\
	  }\href@noop {} {\bibfield  {journal} {\bibinfo  {journal} {Phys. Rev. B}\
	  }\textbf {\bibinfo {volume} {89}},\ \bibinfo {pages} {045422} (\bibinfo
	  {year} {2014})}\BibitemShut {NoStop}%
	\bibitem [{\citenamefont {He}\ \emph {et~al.}(2014)\citenamefont {He},
	  \citenamefont {Wu}, \citenamefont {Choy}, \citenamefont {Liu}, \citenamefont
	  {Tanaka},\ and\ \citenamefont {Law}}]{HeNatureC2014a}%
	  \BibitemOpen
	  \bibfield  {author} {\bibinfo {author} {\bibfnamefont {J.~J.}\ \bibnamefont
	  {He}}, \bibinfo {author} {\bibfnamefont {J.}~\bibnamefont {Wu}}, \bibinfo
	  {author} {\bibfnamefont {T.-P.}\ \bibnamefont {Choy}}, \bibinfo {author}
	  {\bibfnamefont {X.-J.}\ \bibnamefont {Liu}}, \bibinfo {author} {\bibfnamefont
	  {Y.}~\bibnamefont {Tanaka}}, \ and\ \bibinfo {author} {\bibfnamefont {K.~T.}\
	  \bibnamefont {Law}},\ }\href@noop {} {\bibfield  {journal} {\bibinfo
	  {journal} {Nature Commun.}\ }\textbf {\bibinfo {volume} {5}},\ \bibinfo
	  {pages} {3232} (\bibinfo {year} {2014})}\BibitemShut {NoStop}%
	\bibitem [{\citenamefont {Ishizaka}\ \emph {et~al.}(1995)\citenamefont
	  {Ishizaka}, \citenamefont {Sone},\ and\ \citenamefont
	  {Ando}}]{IshizakaPRB1995a}%
	  \BibitemOpen
	  \bibfield  {author} {\bibinfo {author} {\bibfnamefont {S.}~\bibnamefont
	  {Ishizaka}}, \bibinfo {author} {\bibfnamefont {J.}~\bibnamefont {Sone}}, \
	  and\ \bibinfo {author} {\bibfnamefont {T.}~\bibnamefont {Ando}},\ }\href@noop
	  {} {\bibfield  {journal} {\bibinfo  {journal} {Phys. Rev. B}\ }\textbf
	  {\bibinfo {volume} {52}},\ \bibinfo {pages} {8358} (\bibinfo {year}
	  {1995})}\BibitemShut {NoStop}%
	\bibitem [{\citenamefont {Deacon}\ \emph {et~al.}(2015)\citenamefont {Deacon},
	  \citenamefont {Oiwa}, \citenamefont {Sailer}, \citenamefont {Baba},
	  \citenamefont {Kanai}, \citenamefont {Shibata}, \citenamefont {Hirakawa},\
	  and\ \citenamefont {Tarucha}}]{DeaconNatureC2015a}%
	  \BibitemOpen
	  \bibfield  {author} {\bibinfo {author} {\bibfnamefont {R.~S.}\ \bibnamefont
	  {Deacon}}, \bibinfo {author} {\bibfnamefont {A.}~\bibnamefont {Oiwa}},
	  \bibinfo {author} {\bibfnamefont {J.}~\bibnamefont {Sailer}}, \bibinfo
	  {author} {\bibfnamefont {S.}~\bibnamefont {Baba}}, \bibinfo {author}
	  {\bibfnamefont {Y.}~\bibnamefont {Kanai}}, \bibinfo {author} {\bibfnamefont
	  {K.}~\bibnamefont {Shibata}}, \bibinfo {author} {\bibfnamefont
	  {K.}~\bibnamefont {Hirakawa}}, \ and\ \bibinfo {author} {\bibfnamefont
	  {S.}~\bibnamefont {Tarucha}},\ }\href@noop {} {\bibfield  {journal} {\bibinfo
	   {journal} {Nature Commun.}\ }\textbf {\bibinfo {volume} {6}},\ \bibinfo
	  {pages} {7446} (\bibinfo {year} {2015})}\BibitemShut {NoStop}%
	\bibitem [{\citenamefont {Recher}\ \emph {et~al.}(2001)\citenamefont {Recher},
	  \citenamefont {Sukhorukov},\ and\ \citenamefont {Loss}}]{RecherPRB2001a}%
	  \BibitemOpen
	  \bibfield  {author} {\bibinfo {author} {\bibfnamefont {P.}~\bibnamefont
	  {Recher}}, \bibinfo {author} {\bibfnamefont {E.~V.}\ \bibnamefont
	  {Sukhorukov}}, \ and\ \bibinfo {author} {\bibfnamefont {D.}~\bibnamefont
	  {Loss}},\ }\href@noop {} {\bibfield  {journal} {\bibinfo  {journal} {Phys.
	  Rev. B}\ }\textbf {\bibinfo {volume} {63}},\ \bibinfo {pages} {165314}
	  (\bibinfo {year} {2001})}\BibitemShut {NoStop}%
	\bibitem [{\citenamefont {Sothmann}\ \emph {et~al.}(2014)\citenamefont
	  {Sothmann}, \citenamefont {Weiss}, \citenamefont {Governale},\ and\
	  \citenamefont {K\"onig}}]{SothmannPRB2014a}%
	  \BibitemOpen
	  \bibfield  {author} {\bibinfo {author} {\bibfnamefont {B.}~\bibnamefont
	  {Sothmann}}, \bibinfo {author} {\bibfnamefont {S.}~\bibnamefont {Weiss}},
	  \bibinfo {author} {\bibfnamefont {M.}~\bibnamefont {Governale}}, \ and\
	  \bibinfo {author} {\bibfnamefont {J.}~\bibnamefont {K\"onig}},\ }\href@noop
	  {} {\bibfield  {journal} {\bibinfo  {journal} {Phys. Rev. B}\ }\textbf
	  {\bibinfo {volume} {90}},\ \bibinfo {pages} {220501} (\bibinfo {year}
	  {2014})}\BibitemShut {NoStop}%
	\bibitem [{\citenamefont {Hammer}\ \emph {et~al.}(2007)\citenamefont {Hammer},
	  \citenamefont {Cuevas}, \citenamefont {Bergeret},\ and\ \citenamefont
	  {Belzig}}]{HammerPRB2007a}%
	  \BibitemOpen
	  \bibfield  {author} {\bibinfo {author} {\bibfnamefont {J.~C.}\ \bibnamefont
	  {Hammer}}, \bibinfo {author} {\bibfnamefont {J.~C.}\ \bibnamefont {Cuevas}},
	  \bibinfo {author} {\bibfnamefont {F.~S.}\ \bibnamefont {Bergeret}}, \ and\
	  \bibinfo {author} {\bibfnamefont {W.}~\bibnamefont {Belzig}},\ }\href@noop {}
	  {\bibfield  {journal} {\bibinfo  {journal} {Phys. Rev. B}\ }\textbf {\bibinfo
	  {volume} {76}},\ \bibinfo {pages} {064514} (\bibinfo {year}
	  {2007})}\BibitemShut {NoStop}%
	\bibitem [{\citenamefont {Chevallier}\ \emph {et~al.}(2011)\citenamefont
	  {Chevallier}, \citenamefont {Rech}, \citenamefont {Jonckheere},\ and\
	  \citenamefont {Martin}}]{ChevallierPRB2011a}%
	  \BibitemOpen
	  \bibfield  {author} {\bibinfo {author} {\bibfnamefont {D.}~\bibnamefont
	  {Chevallier}}, \bibinfo {author} {\bibfnamefont {J.}~\bibnamefont {Rech}},
	  \bibinfo {author} {\bibfnamefont {T.}~\bibnamefont {Jonckheere}}, \ and\
	  \bibinfo {author} {\bibfnamefont {T.}~\bibnamefont {Martin}},\ }\href@noop {}
	  {\bibfield  {journal} {\bibinfo  {journal} {Phys. Rev. B}\ }\textbf {\bibinfo
	  {volume} {83}},\ \bibinfo {pages} {125421} (\bibinfo {year}
	  {2011})}\BibitemShut {NoStop}%
	\bibitem [{\citenamefont {Rech}\ \emph {et~al.}(2012)\citenamefont {Rech},
	  \citenamefont {Chevallier}, \citenamefont {Jonckheere},\ and\ \citenamefont
	  {Martin}}]{RechPRB2012a}%
	  \BibitemOpen
	  \bibfield  {author} {\bibinfo {author} {\bibfnamefont {J.}~\bibnamefont
	  {Rech}}, \bibinfo {author} {\bibfnamefont {D.}~\bibnamefont {Chevallier}},
	  \bibinfo {author} {\bibfnamefont {T.}~\bibnamefont {Jonckheere}}, \ and\
	  \bibinfo {author} {\bibfnamefont {T.}~\bibnamefont {Martin}},\ }\href@noop {}
	  {\bibfield  {journal} {\bibinfo  {journal} {Phys. Rev. B}\ }\textbf {\bibinfo
	  {volume} {85}},\ \bibinfo {pages} {035419} (\bibinfo {year}
	  {2012})}\BibitemShut {NoStop}%
	\bibitem [{\citenamefont {Droste}\ \emph {et~al.}(2012)\citenamefont {Droste},
	  \citenamefont {Andergassen},\ and\ \citenamefont
	  {Splettstoesser}}]{DrosteJP2012a}%
	  \BibitemOpen
	  \bibfield  {author} {\bibinfo {author} {\bibfnamefont {S.}~\bibnamefont
	  {Droste}}, \bibinfo {author} {\bibfnamefont {S.}~\bibnamefont {Andergassen}},
	  \ and\ \bibinfo {author} {\bibfnamefont {J.}~\bibnamefont {Splettstoesser}},\
	  }\href@noop {} {\bibfield  {journal} {\bibinfo  {journal} {J. Phys.: Condens.
	  Matter}\ }\textbf {\bibinfo {volume} {24}},\ \bibinfo {pages} {415301}
	  (\bibinfo {year} {2012})}\BibitemShut {NoStop}%
	\bibitem [{\citenamefont {Trocha}\ and\ \citenamefont
	  {Weymann}(2015)}]{TrochaPRB2015a}%
	  \BibitemOpen
	  \bibfield  {author} {\bibinfo {author} {\bibfnamefont {P.}~\bibnamefont
	  {Trocha}}\ and\ \bibinfo {author} {\bibfnamefont {I.}~\bibnamefont
	  {Weymann}},\ }\href@noop {} {\bibfield  {journal} {\bibinfo  {journal} {Phys.
	  Rev. B}\ }\textbf {\bibinfo {volume} {91}},\ \bibinfo {pages} {235424}
	  (\bibinfo {year} {2015})}\BibitemShut {NoStop}%
	\bibitem [{\citenamefont {Belzig}\ and\ \citenamefont
	  {Samuelsson}(2003)}]{BelzigEPL2003a}%
	  \BibitemOpen
	  \bibfield  {author} {\bibinfo {author} {\bibfnamefont {W.}~\bibnamefont
	  {Belzig}}\ and\ \bibinfo {author} {\bibfnamefont {P.}~\bibnamefont
	  {Samuelsson}},\ }\href@noop {} {\bibfield  {journal} {\bibinfo  {journal}
	  {Europhys. Lett.}\ }\textbf {\bibinfo {volume} {64}},\ \bibinfo {pages} {253}
	  (\bibinfo {year} {2003})}\BibitemShut {NoStop}%
	\bibitem [{\citenamefont {Governale}\ \emph {et~al.}(2008)\citenamefont
	  {Governale}, \citenamefont {Pala},\ and\ \citenamefont
	  {K\"onig}}]{GovernalePRB2008a}%
	  \BibitemOpen
	  \bibfield  {author} {\bibinfo {author} {\bibfnamefont {M.}~\bibnamefont
	  {Governale}}, \bibinfo {author} {\bibfnamefont {M.~G.}\ \bibnamefont {Pala}},
	  \ and\ \bibinfo {author} {\bibfnamefont {J.}~\bibnamefont {K\"onig}},\
	  }\href@noop {} {\bibfield  {journal} {\bibinfo  {journal} {Phys. Rev. B}\
	  }\textbf {\bibinfo {volume} {77}},\ \bibinfo {pages} {134513} (\bibinfo
	  {year} {2008})}\BibitemShut {NoStop}%
	\bibitem [{\citenamefont {Morten}\ \emph {et~al.}(2008)\citenamefont {Morten},
	  \citenamefont {Huertas-Hernando}, \citenamefont {Belzig},\ and\ \citenamefont
	  {Brataas}}]{MortenPRB2008a}%
	  \BibitemOpen
	  \bibfield  {author} {\bibinfo {author} {\bibfnamefont {J.~P.}\ \bibnamefont
	  {Morten}}, \bibinfo {author} {\bibfnamefont {D.}~\bibnamefont
	  {Huertas-Hernando}}, \bibinfo {author} {\bibfnamefont {W.}~\bibnamefont
	  {Belzig}}, \ and\ \bibinfo {author} {\bibfnamefont {A.}~\bibnamefont
	  {Brataas}},\ }\href@noop {} {\bibfield  {journal} {\bibinfo  {journal} {Phys.
	  Rev. B}\ }\textbf {\bibinfo {volume} {78}},\ \bibinfo {pages} {224515}
	  (\bibinfo {year} {2008})}\BibitemShut {NoStop}%
	\bibitem [{\citenamefont {Futterer}\ \emph {et~al.}(2013)\citenamefont
	  {Futterer}, \citenamefont {Swiebodzinski}, \citenamefont {Governale},\ and\
	  \citenamefont {K\"onig}}]{FuttererPRB2013a}%
	  \BibitemOpen
	  \bibfield  {author} {\bibinfo {author} {\bibfnamefont {D.}~\bibnamefont
	  {Futterer}}, \bibinfo {author} {\bibfnamefont {J.}~\bibnamefont
	  {Swiebodzinski}}, \bibinfo {author} {\bibfnamefont {M.}~\bibnamefont
	  {Governale}}, \ and\ \bibinfo {author} {\bibfnamefont {J.}~\bibnamefont
	  {K\"onig}},\ }\href@noop {} {\bibfield  {journal} {\bibinfo  {journal} {Phys.
	  Rev. B}\ }\textbf {\bibinfo {volume} {87}},\ \bibinfo {pages} {014509}
	  (\bibinfo {year} {2013})}\BibitemShut {NoStop}%
	\bibitem [{\citenamefont {Soller}\ and\ \citenamefont
	  {Komnik}(2014)}]{SollerEPL2014a}%
	  \BibitemOpen
	  \bibfield  {author} {\bibinfo {author} {\bibfnamefont {H.}~\bibnamefont
	  {Soller}}\ and\ \bibinfo {author} {\bibfnamefont {A.}~\bibnamefont
	  {Komnik}},\ }\href@noop {} {\bibfield  {journal} {\bibinfo  {journal}
	  {Europhys. Lett.}\ }\textbf {\bibinfo {volume} {106}},\ \bibinfo {pages}
	  {37009} (\bibinfo {year} {2014})}\BibitemShut {NoStop}%
	\bibitem [{\citenamefont {{Stegmann}}\ and\ \citenamefont
	  {{K{\"o}nig}}()}]{StegmannArXiv2016a}%
	  \BibitemOpen
	  \bibfield  {author} {\bibinfo {author} {\bibfnamefont {P.}~\bibnamefont
	  {{Stegmann}}}\ and\ \bibinfo {author} {\bibfnamefont {J.}~\bibnamefont
	  {{K{\"o}nig}}},\ }\href@noop {} {\bibinfo  {journal} {arXiv:1605.09258v1}\
	  }\BibitemShut {NoStop}%
	\bibitem [{\citenamefont {F\"ul\"op}\ \emph {et~al.}(2015)\citenamefont
	  {F\"ul\"op}, \citenamefont {Dom\'{\i}nguez}, \citenamefont {d'Hollosy},
	  \citenamefont {Baumgartner}, \citenamefont {Makk}, \citenamefont {Madsen},
	  \citenamefont {Guzenko}, \citenamefont {Nyg\aa{}rd}, \citenamefont
	  {Sch\"onenberger}, \citenamefont {Levy~Yeyati},\ and\ \citenamefont
	  {Csonka}}]{FueloepPRL2015a}%
	  \BibitemOpen
	\bibfield  {journal} {  }\bibfield  {author} {\bibinfo {author} {\bibfnamefont
	  {G.}~\bibnamefont {F\"ul\"op}}, \bibinfo {author} {\bibfnamefont
	  {F.}~\bibnamefont {Dom\'{\i}nguez}}, \bibinfo {author} {\bibfnamefont
	  {S.}~\bibnamefont {d'Hollosy}}, \bibinfo {author} {\bibfnamefont
	  {A.}~\bibnamefont {Baumgartner}}, \bibinfo {author} {\bibfnamefont
	  {P.}~\bibnamefont {Makk}}, \bibinfo {author} {\bibfnamefont {M.~H.}\
	  \bibnamefont {Madsen}}, \bibinfo {author} {\bibfnamefont {V.~A.}\
	  \bibnamefont {Guzenko}}, \bibinfo {author} {\bibfnamefont {J.}~\bibnamefont
	  {Nyg\aa{}rd}}, \bibinfo {author} {\bibfnamefont {C.}~\bibnamefont
	  {Sch\"onenberger}}, \bibinfo {author} {\bibfnamefont {A.}~\bibnamefont
	  {Levy~Yeyati}}, \ and\ \bibinfo {author} {\bibfnamefont {S.}~\bibnamefont
	  {Csonka}},\ }\href@noop {} {\bibfield  {journal} {\bibinfo  {journal} {Phys.
	  Rev. Lett.}\ }\textbf {\bibinfo {volume} {115}},\ \bibinfo {pages} {227003}
	  (\bibinfo {year} {2015})}\BibitemShut {NoStop}%
	\bibitem [{\citenamefont {Machon}\ \emph {et~al.}(2013)\citenamefont {Machon},
	  \citenamefont {Eschrig},\ and\ \citenamefont {Belzig}}]{MachonPRL2013a}%
	  \BibitemOpen
	  \bibfield  {author} {\bibinfo {author} {\bibfnamefont {P.}~\bibnamefont
	  {Machon}}, \bibinfo {author} {\bibfnamefont {M.}~\bibnamefont {Eschrig}}, \
	  and\ \bibinfo {author} {\bibfnamefont {W.}~\bibnamefont {Belzig}},\
	  }\href@noop {} {\bibfield  {journal} {\bibinfo  {journal} {Phys. Rev. Lett.}\
	  }\textbf {\bibinfo {volume} {110}},\ \bibinfo {pages} {047002} (\bibinfo
	  {year} {2013})}\BibitemShut {NoStop}%
	\bibitem [{\citenamefont {Cao}\ \emph {et~al.}(2015)\citenamefont {Cao},
	  \citenamefont {Fang}, \citenamefont {Li},\ and\ \citenamefont
	  {Luo}}]{CaoAPL2015a}%
	  \BibitemOpen
	  \bibfield  {author} {\bibinfo {author} {\bibfnamefont {Z.}~\bibnamefont
	  {Cao}}, \bibinfo {author} {\bibfnamefont {T.-F.}\ \bibnamefont {Fang}},
	  \bibinfo {author} {\bibfnamefont {L.}~\bibnamefont {Li}}, \ and\ \bibinfo
	  {author} {\bibfnamefont {H.-G.}\ \bibnamefont {Luo}},\ }\href@noop {}
	  {\bibfield  {journal} {\bibinfo  {journal} {Appl. Phys. Lett.}\ }\textbf
	  {\bibinfo {volume} {107}},\ \bibinfo {eid} {212601} (\bibinfo {year}
	  {2015})}\BibitemShut {NoStop}%
	\bibitem [{\citenamefont {Hofstetter}\ \emph {et~al.}(2009)\citenamefont
	  {Hofstetter}, \citenamefont {Csonka}, \citenamefont {Nygard},\ and\
	  \citenamefont {Schonenberger}}]{HofstetterNature2009a}%
	  \BibitemOpen
	  \bibfield  {author} {\bibinfo {author} {\bibfnamefont {L.}~\bibnamefont
	  {Hofstetter}}, \bibinfo {author} {\bibfnamefont {S.}~\bibnamefont {Csonka}},
	  \bibinfo {author} {\bibfnamefont {J.}~\bibnamefont {Nygard}}, \ and\ \bibinfo
	  {author} {\bibfnamefont {C.}~\bibnamefont {Schonenberger}},\ }\href@noop {}
	  {\bibfield  {journal} {\bibinfo  {journal} {Nature}\ }\textbf {\bibinfo
	  {volume} {461}},\ \bibinfo {pages} {960} (\bibinfo {year}
	  {2009})}\BibitemShut {NoStop}%
	\bibitem [{\citenamefont {Beckmann}\ \emph {et~al.}(2004)\citenamefont
	  {Beckmann}, \citenamefont {Weber},\ and\ \citenamefont
	  {v.~L\"ohneysen}}]{BeckmannPRL2004a}%
	  \BibitemOpen
	  \bibfield  {author} {\bibinfo {author} {\bibfnamefont {D.}~\bibnamefont
	  {Beckmann}}, \bibinfo {author} {\bibfnamefont {H.~B.}\ \bibnamefont {Weber}},
	  \ and\ \bibinfo {author} {\bibfnamefont {H.}~\bibnamefont {v.~L\"ohneysen}},\
	  }\href@noop {} {\bibfield  {journal} {\bibinfo  {journal} {Phys. Rev. Lett.}\
	  }\textbf {\bibinfo {volume} {93}},\ \bibinfo {pages} {197003} (\bibinfo
	  {year} {2004})}\BibitemShut {NoStop}%
	\bibitem [{\citenamefont {Samm}\ \emph {et~al.}(2014)\citenamefont {Samm},
	  \citenamefont {Gramich}, \citenamefont {Baumgartner}, \citenamefont {Weiss},\
	  and\ \citenamefont {Sch\"onenberger}}]{SammJAP2014a}%
	  \BibitemOpen
	  \bibfield  {author} {\bibinfo {author} {\bibfnamefont {J.}~\bibnamefont
	  {Samm}}, \bibinfo {author} {\bibfnamefont {J.}~\bibnamefont {Gramich}},
	  \bibinfo {author} {\bibfnamefont {A.}~\bibnamefont {Baumgartner}}, \bibinfo
	  {author} {\bibfnamefont {M.}~\bibnamefont {Weiss}}, \ and\ \bibinfo {author}
	  {\bibfnamefont {C.}~\bibnamefont {Sch\"onenberger}},\ }\href@noop {}
	  {\bibfield  {journal} {\bibinfo  {journal} {J. App. Phys.}\ }\textbf
	  {\bibinfo {volume} {115}},\ \bibinfo {eid} {174309} (\bibinfo {year}
	  {2014})}\BibitemShut {NoStop}%
	\bibitem [{\citenamefont {Eldridge}\ \emph {et~al.}(2010)\citenamefont
	  {Eldridge}, \citenamefont {Pala}, \citenamefont {Governale},\ and\
	  \citenamefont {K\"onig}}]{EldridgePRB2010a}%
	  \BibitemOpen
	  \bibfield  {author} {\bibinfo {author} {\bibfnamefont {J.}~\bibnamefont
	  {Eldridge}}, \bibinfo {author} {\bibfnamefont {M.~G.}\ \bibnamefont {Pala}},
	  \bibinfo {author} {\bibfnamefont {M.}~\bibnamefont {Governale}}, \ and\
	  \bibinfo {author} {\bibfnamefont {J.}~\bibnamefont {K\"onig}},\ }\href@noop
	  {} {\bibfield  {journal} {\bibinfo  {journal} {Phys. Rev. B}\ }\textbf
	  {\bibinfo {volume} {82}},\ \bibinfo {pages} {184507} (\bibinfo {year}
	  {2010})}\BibitemShut {NoStop}%
	\bibitem [{Note1()}]{Note1}%
	  \BibitemOpen
	  \bibinfo {note} {Due to the absence of an applied magnetic field it is
	  possible to choose the spin-quantization axis such that the interdot
	  tunneling in the presence of the SO coupling is diagonal in the spin
	  space.}\BibitemShut {Stop}%
	\bibitem [{\citenamefont {Fasth}\ \emph {et~al.}(2007)\citenamefont {Fasth},
	  \citenamefont {Fuhrer}, \citenamefont {Samuelson}, \citenamefont {Golovach},\
	  and\ \citenamefont {Loss}}]{FasthPRL2007a}%
	  \BibitemOpen
	  \bibfield  {author} {\bibinfo {author} {\bibfnamefont {C.}~\bibnamefont
	  {Fasth}}, \bibinfo {author} {\bibfnamefont {A.}~\bibnamefont {Fuhrer}},
	  \bibinfo {author} {\bibfnamefont {L.}~\bibnamefont {Samuelson}}, \bibinfo
	  {author} {\bibfnamefont {V.~N.}\ \bibnamefont {Golovach}}, \ and\ \bibinfo
	  {author} {\bibfnamefont {D.}~\bibnamefont {Loss}},\ }\href@noop {} {\bibfield
	   {journal} {\bibinfo  {journal} {Phys. Rev. Lett.}\ }\textbf {\bibinfo
	  {volume} {98}},\ \bibinfo {pages} {266801} (\bibinfo {year}
	  {2007})}\BibitemShut {NoStop}%
	\bibitem [{\citenamefont {Est\'evez~Hern\'andez}\ \emph
	  {et~al.}(2010)\citenamefont {Est\'evez~Hern\'andez}, \citenamefont {Akabori},
	  \citenamefont {Sladek}, \citenamefont {Volk}, \citenamefont {Alagha},
	  \citenamefont {Hardtdegen}, \citenamefont {Pala}, \citenamefont {Demarina},
	  \citenamefont {Gr\"utzmacher},\ and\ \citenamefont
	  {Sch\"apers}}]{HernandezPRB2010a}%
	  \BibitemOpen
	  \bibfield  {author} {\bibinfo {author} {\bibfnamefont {S.}~\bibnamefont
	  {Est\'evez~Hern\'andez}}, \bibinfo {author} {\bibfnamefont {M.}~\bibnamefont
	  {Akabori}}, \bibinfo {author} {\bibfnamefont {K.}~\bibnamefont {Sladek}},
	  \bibinfo {author} {\bibfnamefont {C.}~\bibnamefont {Volk}}, \bibinfo {author}
	  {\bibfnamefont {S.}~\bibnamefont {Alagha}}, \bibinfo {author} {\bibfnamefont
	  {H.}~\bibnamefont {Hardtdegen}}, \bibinfo {author} {\bibfnamefont {M.~G.}\
	  \bibnamefont {Pala}}, \bibinfo {author} {\bibfnamefont {N.}~\bibnamefont
	  {Demarina}}, \bibinfo {author} {\bibfnamefont {D.}~\bibnamefont
	  {Gr\"utzmacher}}, \ and\ \bibinfo {author} {\bibfnamefont {T.}~\bibnamefont
	  {Sch\"apers}},\ }\href@noop {} {\bibfield  {journal} {\bibinfo  {journal}
	  {Phys. Rev. B}\ }\textbf {\bibinfo {volume} {82}},\ \bibinfo {pages} {235303}
	  (\bibinfo {year} {2010})}\BibitemShut {NoStop}%
	\bibitem [{\citenamefont {Nadj-Perge}\ \emph {et~al.}(2012)\citenamefont
	  {Nadj-Perge}, \citenamefont {Pribiag}, \citenamefont {van~den Berg},
	  \citenamefont {Zuo}, \citenamefont {Plissard}, \citenamefont {Bakkers},
	  \citenamefont {Frolov},\ and\ \citenamefont
	  {Kouwenhoven}}]{Nadj-PergePRL2012a}%
	  \BibitemOpen
	  \bibfield  {author} {\bibinfo {author} {\bibfnamefont {S.}~\bibnamefont
	  {Nadj-Perge}}, \bibinfo {author} {\bibfnamefont {V.~S.}\ \bibnamefont
	  {Pribiag}}, \bibinfo {author} {\bibfnamefont {J.~W.~G.}\ \bibnamefont
	  {van~den Berg}}, \bibinfo {author} {\bibfnamefont {K.}~\bibnamefont {Zuo}},
	  \bibinfo {author} {\bibfnamefont {S.~R.}\ \bibnamefont {Plissard}}, \bibinfo
	  {author} {\bibfnamefont {E.~P. A.~M.}\ \bibnamefont {Bakkers}}, \bibinfo
	  {author} {\bibfnamefont {S.~M.}\ \bibnamefont {Frolov}}, \ and\ \bibinfo
	  {author} {\bibfnamefont {L.~P.}\ \bibnamefont {Kouwenhoven}},\ }\href@noop {}
	  {\bibfield  {journal} {\bibinfo  {journal} {Phys. Rev. Lett.}\ }\textbf
	  {\bibinfo {volume} {108}},\ \bibinfo {pages} {166801} (\bibinfo {year}
	  {2012})}\BibitemShut {NoStop}%
	\bibitem [{\citenamefont {van Weperen}\ \emph {et~al.}(2015)\citenamefont {van
	  Weperen}, \citenamefont {Tarasinski}, \citenamefont {Eeltink}, \citenamefont
	  {Pribiag}, \citenamefont {Plissard}, \citenamefont {Bakkers}, \citenamefont
	  {Kouwenhoven},\ and\ \citenamefont {Wimmer}}]{WeperenPRB2015a}%
	  \BibitemOpen
	  \bibfield  {author} {\bibinfo {author} {\bibfnamefont {I.}~\bibnamefont {van
	  Weperen}}, \bibinfo {author} {\bibfnamefont {B.}~\bibnamefont {Tarasinski}},
	  \bibinfo {author} {\bibfnamefont {D.}~\bibnamefont {Eeltink}}, \bibinfo
	  {author} {\bibfnamefont {V.~S.}\ \bibnamefont {Pribiag}}, \bibinfo {author}
	  {\bibfnamefont {S.~R.}\ \bibnamefont {Plissard}}, \bibinfo {author}
	  {\bibfnamefont {E.~P. A.~M.}\ \bibnamefont {Bakkers}}, \bibinfo {author}
	  {\bibfnamefont {L.~P.}\ \bibnamefont {Kouwenhoven}}, \ and\ \bibinfo {author}
	  {\bibfnamefont {M.}~\bibnamefont {Wimmer}},\ }\href@noop {} {\bibfield
	  {journal} {\bibinfo  {journal} {Phys. Rev. B}\ }\textbf {\bibinfo {volume}
	  {91}},\ \bibinfo {pages} {201413} (\bibinfo {year} {2015})}\BibitemShut
	  {NoStop}%
	\bibitem [{\citenamefont {Rozhkov}\ and\ \citenamefont
	  {Arovas}(2000)}]{RozhkovPRB2000a}%
	  \BibitemOpen
	  \bibfield  {author} {\bibinfo {author} {\bibfnamefont {A.~V.}\ \bibnamefont
	  {Rozhkov}}\ and\ \bibinfo {author} {\bibfnamefont {D.~P.}\ \bibnamefont
	  {Arovas}},\ }\href@noop {} {\bibfield  {journal} {\bibinfo  {journal} {Phys.
	  Rev. B}\ }\textbf {\bibinfo {volume} {62}},\ \bibinfo {pages} {6687}
	  (\bibinfo {year} {2000})}\BibitemShut {NoStop}%
	\bibitem [{\citenamefont {Meng}\ \emph {et~al.}(2009)\citenamefont {Meng},
	  \citenamefont {Florens},\ and\ \citenamefont {Simon}}]{MengPRB2009a}%
	  \BibitemOpen
	  \bibfield  {author} {\bibinfo {author} {\bibfnamefont {T.}~\bibnamefont
	  {Meng}}, \bibinfo {author} {\bibfnamefont {S.}~\bibnamefont {Florens}}, \
	  and\ \bibinfo {author} {\bibfnamefont {P.}~\bibnamefont {Simon}},\
	  }\href@noop {} {\bibfield  {journal} {\bibinfo  {journal} {Phys. Rev. B}\
	  }\textbf {\bibinfo {volume} {79}},\ \bibinfo {pages} {224521} (\bibinfo
	  {year} {2009})}\BibitemShut {NoStop}%
	\bibitem [{\citenamefont {Rajabi}\ \emph {et~al.}(2013)\citenamefont {Rajabi},
	  \citenamefont {P\"oltl},\ and\ \citenamefont {Governale}}]{RajabiPRL2013a}%
	  \BibitemOpen
	  \bibfield  {author} {\bibinfo {author} {\bibfnamefont {L.}~\bibnamefont
	  {Rajabi}}, \bibinfo {author} {\bibfnamefont {C.}~\bibnamefont {P\"oltl}}, \
	  and\ \bibinfo {author} {\bibfnamefont {M.}~\bibnamefont {Governale}},\
	  }\href@noop {} {\bibfield  {journal} {\bibinfo  {journal} {Phys. Rev. Lett.}\
	  }\textbf {\bibinfo {volume} {111}},\ \bibinfo {pages} {067002} (\bibinfo
	  {year} {2013})}\BibitemShut {NoStop}%
	\bibitem [{\citenamefont {Braggio}\ \emph {et~al.}(2011)\citenamefont
	  {Braggio}, \citenamefont {Governale}, \citenamefont {Pala},\ and\
	  \citenamefont {K\"onig}}]{BraggioSSC2011a}%
	  \BibitemOpen
	  \bibfield  {author} {\bibinfo {author} {\bibfnamefont {A.}~\bibnamefont
	  {Braggio}}, \bibinfo {author} {\bibfnamefont {M.}~\bibnamefont {Governale}},
	  \bibinfo {author} {\bibfnamefont {M.~G.}\ \bibnamefont {Pala}}, \ and\
	  \bibinfo {author} {\bibfnamefont {J.}~\bibnamefont {K\"onig}},\ }\href@noop
	  {} {\bibfield  {journal} {\bibinfo  {journal} {Solid State Commun.}\ }\textbf
	  {\bibinfo {volume} {151}},\ \bibinfo {pages} {155} (\bibinfo {year}
	  {2011})}\BibitemShut {NoStop}%
	\bibitem [{\citenamefont {Amitai}\ \emph {et~al.}(2016)\citenamefont {Amitai},
	  \citenamefont {Tiwari}, \citenamefont {Walter}, \citenamefont {Schmidt},\
	  and\ \citenamefont {Nigg}}]{AmitaiPRB2016a}%
	  \BibitemOpen
	  \bibfield  {author} {\bibinfo {author} {\bibfnamefont {E.}~\bibnamefont
	  {Amitai}}, \bibinfo {author} {\bibfnamefont {R.~P.}\ \bibnamefont {Tiwari}},
	  \bibinfo {author} {\bibfnamefont {S.}~\bibnamefont {Walter}}, \bibinfo
	  {author} {\bibfnamefont {T.~L.}\ \bibnamefont {Schmidt}}, \ and\ \bibinfo
	  {author} {\bibfnamefont {S.~E.}\ \bibnamefont {Nigg}},\ }\href@noop {}
	  {\bibfield  {journal} {\bibinfo  {journal} {Phys. Rev. B}\ }\textbf {\bibinfo
	  {volume} {93}},\ \bibinfo {pages} {075421} (\bibinfo {year}
	  {2016})}\BibitemShut {NoStop}%
	\bibitem [{\citenamefont {Bagrets}\ and\ \citenamefont {{Y}u.
	  V.~Nazarov}(2003)}]{BagretsPRB2003a}%
	  \BibitemOpen
	  \bibfield  {author} {\bibinfo {author} {\bibfnamefont {D.~A.}\ \bibnamefont
	  {Bagrets}}\ and\ \bibinfo {author} {\bibnamefont {{Y}u. V.~Nazarov}},\
	  }\href@noop {} {\bibfield  {journal} {\bibinfo  {journal} {Phys. Rev. B}\
	  }\textbf {\bibinfo {volume} {67}},\ \bibinfo {pages} {085316} (\bibinfo
	  {year} {2003})}\BibitemShut {NoStop}%
	\bibitem [{\citenamefont {Flindt}\ \emph {et~al.}(2004)\citenamefont {Flindt},
	  \citenamefont {Novotn\'y},\ and\ \citenamefont {Jauho}}]{FlindtPRB2004a}%
	  \BibitemOpen
	  \bibfield  {author} {\bibinfo {author} {\bibfnamefont {C.}~\bibnamefont
	  {Flindt}}, \bibinfo {author} {\bibfnamefont {T.}\ \bibnamefont
	  {Novotn\'y}}, \ and\ \bibinfo {author} {\bibfnamefont {A.-P.}\ \bibnamefont
	  {Jauho}},\ }\href@noop {} {\bibfield  {journal} {\bibinfo  {journal} {Phys.
	  Rev. B}\ }\textbf {\bibinfo {volume} {70}},\ \bibinfo {pages} {205334}
	  (\bibinfo {year} {2004})}\BibitemShut {NoStop}%
	\bibitem [{\citenamefont {Braggio}\ \emph {et~al.}(2006)\citenamefont
	  {Braggio}, \citenamefont {K\"onig},\ and\ \citenamefont
	  {Fazio}}]{BraggioPRL2006a}%
	  \BibitemOpen
	  \bibfield  {author} {\bibinfo {author} {\bibfnamefont {A.}~\bibnamefont
	  {Braggio}}, \bibinfo {author} {\bibfnamefont {J.}~\bibnamefont {K\"onig}}, \
	  and\ \bibinfo {author} {\bibfnamefont {R.}~\bibnamefont {Fazio}},\
	  }\href@noop {} {\bibfield  {journal} {\bibinfo  {journal} {Phys. Rev. Lett.}\
	  }\textbf {\bibinfo {volume} {96}},\ \bibinfo {pages} {026805} (\bibinfo
	  {year} {2006})}\BibitemShut {NoStop}%
	\bibitem [{\citenamefont {Kaiser}\ and\ \citenamefont
	  {Kohler}(2007)}]{KaiserAP2007a}%
	  \BibitemOpen
	  \bibfield  {author} {\bibinfo {author} {\bibfnamefont {F.~J.}\ \bibnamefont
	  {Kaiser}}\ and\ \bibinfo {author} {\bibfnamefont {S.}~\bibnamefont
	  {Kohler}},\ }\href@noop {} {\bibfield  {journal} {\bibinfo  {journal} {Ann.
	  Phys. (Leipzig)}\ }\textbf {\bibinfo {volume} {16}},\ \bibinfo {pages} {702}
	  (\bibinfo {year} {2007})}\BibitemShut {NoStop}%
	\bibitem [{\citenamefont {Hussein}\ and\ \citenamefont
	  {Kohler}(2014)}]{HusseinPRB2014a}%
	  \BibitemOpen
	  \bibfield  {author} {\bibinfo {author} {\bibfnamefont {R.}~\bibnamefont
	  {Hussein}}\ and\ \bibinfo {author} {\bibfnamefont {S.}~\bibnamefont
	  {Kohler}},\ }\href@noop {} {\bibfield  {journal} {\bibinfo  {journal} {Phys.
	  Rev. B}\ }\textbf {\bibinfo {volume} {89}},\ \bibinfo {pages} {205424}
	  (\bibinfo {year} {2014})}\BibitemShut {NoStop}%
	\bibitem [{\citenamefont {Flindt}\ \emph {et~al.}(2008)\citenamefont {Flindt},
	  \citenamefont {Novotn\'y}, \citenamefont {Braggio}, \citenamefont
	  {Sassetti},\ and\ \citenamefont {Jauho}}]{FlindtPRL2008a}%
	  \BibitemOpen
	  \bibfield  {author} {\bibinfo {author} {\bibfnamefont {C.}~\bibnamefont
	  {Flindt}}, \bibinfo {author} {\bibfnamefont {T.}~\bibnamefont {Novotn\'y}},
	  \bibinfo {author} {\bibfnamefont {A.}~\bibnamefont {Braggio}}, \bibinfo
	  {author} {\bibfnamefont {M.}~\bibnamefont {Sassetti}}, \ and\ \bibinfo
	  {author} {\bibfnamefont {A.-P.}\ \bibnamefont {Jauho}},\ }\href@noop {}
	  {\bibfield  {journal} {\bibinfo  {journal} {Phys. Rev. Lett.}\ }\textbf
	  {\bibinfo {volume} {100}},\ \bibinfo {pages} {150601} (\bibinfo {year}
	  {2008})}\BibitemShut {NoStop}%
	\bibitem [{\citenamefont {Flindt}\ \emph {et~al.}(2010)\citenamefont {Flindt},
	  \citenamefont {Novotn\'y}, \citenamefont {Braggio},\ and\ \citenamefont
	  {Jauho}}]{FlindtPRB2010a}%
	  \BibitemOpen
	  \bibfield  {author} {\bibinfo {author} {\bibfnamefont {C.}~\bibnamefont
	  {Flindt}}, \bibinfo {author} {\bibfnamefont {T.}~\bibnamefont {Novotn\'y}},
	  \bibinfo {author} {\bibfnamefont {A.}~\bibnamefont {Braggio}}, \ and\
	  \bibinfo {author} {\bibfnamefont {A.-P.}\ \bibnamefont {Jauho}},\ }\href@noop
	  {} {\bibfield  {journal} {\bibinfo  {journal} {Phys. Rev. B}\ }\textbf
	  {\bibinfo {volume} {82}},\ \bibinfo {pages} {155407} (\bibinfo {year}
	  {2010})}\BibitemShut {NoStop}%
	\bibitem [{Note2()}]{Note2}%
	  \BibitemOpen
	  \bibinfo {note} {Only in the limit $\Gamma _S\to 0$ the hybridized states
	  $\protect \ensuremath {|\pm \delimiter "526930B }$ can be written as linear
	  combination of \protect \ensuremath {|S\delimiter "526930B } and \protect
	  \ensuremath {|d+\delimiter "526930B }.}\BibitemShut {Stop}%
	\bibitem [{\citenamefont {Berestetskii}\ \emph {et~al.}(1982)\citenamefont
	  {Berestetskii}, \citenamefont {Lifshitz},\ and\ \citenamefont
	  {Pitaevskii}}]{Berestetskii1982a}%
	  \BibitemOpen
	  \bibfield  {author} {\bibinfo {author} {\bibfnamefont {V.~B.}\ \bibnamefont
	  {Berestetskii}}, \bibinfo {author} {\bibfnamefont {E.~M.}\ \bibnamefont
	  {Lifshitz}}, \ and\ \bibinfo {author} {\bibfnamefont {L.~P.}\ \bibnamefont
	  {Pitaevskii}},\ }\href@noop {} {\emph {\bibinfo {title} {Quantum
	  Electrodynamics}}},\ \bibinfo {edition} {2nd}\ ed.,\ Vol.~\bibinfo {volume}
	  {4}\ (\bibinfo  {publisher} {Butterworth-Heinemann},\ \bibinfo {year}
	  {1982})\BibitemShut {NoStop}%
	\end{thebibliography}
\end{document}